\documentclass[aps, prb, notitlepage, citeautoscript, superscriptaddress, eqsecnum, reprint]{revtex4-2}

\usepackage{xr}
\usepackage{amsmath}
\usepackage{amsthm}
\usepackage{amsfonts, amssymb,amsxtra}
\usepackage[]{graphicx}
\usepackage{grffile}
\usepackage{amsfonts}
\usepackage{framed}
\usepackage{bbm}
\usepackage{braket}
\usepackage{xcolor}
\usepackage{mathtools}
\usepackage{LatexCommands}
\usepackage{booktabs}

\usepackage{comment}
\usepackage{multirow}

\usepackage[colorlinks=true]{hyperref}
\hypersetup{
    unicode=false,          
    pdftoolbar=true,        
    pdfmenubar=true,        
    pdffitwindow=false,     
    pdfstartview={FitH},    
    pdftitle={},    
    pdfauthor={},     
    pdfsubject={},   
    pdfcreator={},   
    pdfproducer={}, 
    pdfkeywords={} {} {}, 
    pdfnewwindow=true,      
    colorlinks=true,       
    linkcolor=blue, 
    citecolor=blue,        
    filecolor=blue,      
    urlcolor=blue           
}

\begin{document}

\title{
Efficient Magic State Cultivation for $\sqrt{T}$ Gates}

\author{I-Chi Chen}
\email{ichen@lanl.gov}

~\affiliation{Los Alamos National Laboratory, Computing and Artificial Intelligence Division, Los Alamos, NM, USA}

\author{Matheus da Silva Fonseca}
~\affiliation{Los Alamos National Laboratory, Computing and Artificial Intelligence Division, Los Alamos, NM, USA}
~\affiliation{Departamento de Física, Universidade Federal de São Carlos, 13565-905, São Carlos, São Paulo, Brasil }
\email{matheus.sf.mel@gmail.com}

\author{Andrew Sornborger}
~\affiliation{Los Alamos National Laboratory, Computing and Artificial Intelligence Division, Los Alamos, NM, USA}

\begin{abstract}
Recently, experimental and theoretical quantum error correction methodology has seen remarkable breakthroughs. In particular, magic state cultivation has been shown to simplify magic-state preparation and make it feasible for near-term devices. However, recent research on magic state cultivation has focused primarily on the cultivation of $T\ket{+}_L$. Only a few other magic state cultivation methods beyond $T\ket{+}_L$ have been investigated. Here, we generalize phase kickback checks for magic states at arbitrary Clifford hierarchy levels in specific codes. We provide an example of the cultivation of $\sqrt{T}\ket{+}_L$ in the doubled color code and the corresponding escape strategy using lattice surgery from the color code to large rotated surface codes. 
Using state vector simulation for un-grown cultivation, we observe a strong consistency between $S\ket{+}_L$ and $\sqrt{T}\ket{+}_L$ cultivation's performance on the doubled color code. Finally, we discuss the application of the corresponding $\sqrt{T}\ket{+}_L$ cultivation, incorporating the STAR architecture and $T$ gates, for early fault-tolerant quantum computing and its potential to shorten gate synthesis in the fully fault-tolerant quantum computing era. 
\end{abstract}
\date{\today}

\maketitle


\section{Introduction}
Fault-tolerant magic-state preparation plays a key role in a universal fault-tolerant quantum computer~\cite{Shor1996}. Magic State Distillation (MSD)~\cite{Bravyi2005,Litinski2019} is the best-known method for creating a high-fidelity magic state. Although there are experiments on different types of hardware that demonstrate the MSD procedure~\cite{Souza2011,Brown2023,Gupta2024,Sales2025,Hirano2025}, the high space-time overhead remains the main bottleneck to efficiently generate magic states on near-term devices using MSD. An alternative approach to MSD is magic-state preparation using code-switch protocols~\cite{Daguerre25} between, for instance, the $3$D color code supporting a transversal $T$ gate and the $2$D color code, which supports transversal Clifford gates. A recent code switching experiment has demonstrated $5 \times 10^{-4}$ $T\ket{+}_L$ magic state infidelity on a trapped-ion device~\cite{Daguerre2025}. However, this method is restricted to the color code. 

Recently, the development of magic-state cultivation (MSC)~\cite{Gidney2024,Vaknin2025,Chen2025,Sahay2025,Claes2025} promises low space-time overhead and is well-suited for near-term devices with a physical error rate of order $10^{-3}$. Moreover, an MSC experiment~\cite{Rosenfeld2025} has demonstrated low infidelity $\sim 10^{-4}$ magic-state preparation on a quantum superconducting computer for a quantum error correction code. This result sheds light on a viable path to build an early fault-tolerant quantum computer.

The core idea of MSC is to use phase kickback measurements to check the quality of the magic state~\cite{Chamberland2020,Itogawa2025}. The majority of MSC research has focused on using $H_{XY}$ phase kickback checks to achieve higher fidelity for the magic state $T\ket{+}_L$.  Since only self-dual codes support the transversal $H_{XY}$ gate, the transversal $H_{XY}$ check only applies to the $2$D color code~\cite{Gidney2024} and  $\mathrm{SRP}$-$3$ $(5)$ codes~\cite{Chen2025}. At the escape stage, the corresponding code is deformed into a surface code using grafting and lattice surgery. Finally, one discards corrupted syndrome outcomes using soft postselection methods based on the complementary gap. 

In addition to self-dual codes, using the property that the unrotated surface code supports folding transversal $S$ gates~\cite{Moussa2016}, the authors of Ref.~\cite{Sahay2025} were able to apply the $H_{XY}$ check on an unrotated surface code and convert the code back into a rotated surface code after a half round of syndrome extraction~\cite{Chen2024}. At the escape stage, they used unitary growth to expand the code into a larger code. Beyond $T\ket{+}_L$ MSC, using the property that all CSS codes have a transversal $\text{CNOT}$ gate, the authors~\cite{Sahay2025} were able to check $ \ket{CX}_L\equiv \frac{\ket{0,+}_L+\ket{0,-}_L+\ket{1,+}_L}{\sqrt{3}}$, which is the resource state for the logical Toffoli implementation~\cite{Dennis2001}, on the surface code. However, the two methods described above require a $3$-qubit gate, which is not a native gate on most quantum hardware, for the phase kickback check. Furthermore, Ref.~\cite{Liu2026} proposes another MSC for generating $C_{xyz}$ states, which are the magic state resources for generating the $Z$ rotation gate, $\frac{\pi}{6}$. However, the cultivation protocol requires ancilla qutrits, which are not accessible for most quantum hardware, to implement phase kickback measurement.  

In this work, inspired by fold transversal surface code cultivation, which combines control transversal $X$ and control fold transversal $S$ gates as the $H_{XY}$ check on the surface code, we generalize the phase kickback check for the magic state at arbitrary Clifford hierarchy levels.
We take $\sqrt{T}\ket{+}_L$ MSC on the doubled color code as an example and provide the corresponding growth strategy from the doubled color code to a large rotated surface code. We manifest end-to-end simulation for the entire MSC to test the corresponding performance. Finally, we discuss the combination of $\sqrt{T}\ket{+}_L$ and $T\ket{+}_L$ MSC and Space-Time efficient Analog Rotation (STAR) architectures~\cite{Akahoshi2024,Toshio2024,ismail2026transversal,Akahoshi2025} for early fault tolerance and its corresponding advantages over Clifford+$T$ even in the fully fault-tolerant quantum computing era~\cite{Kliuchnikov2023}. 

\section{Methodology\label{sec2}} 
\begin{figure}[t]
\centering
\includegraphics[width=0.9\linewidth]{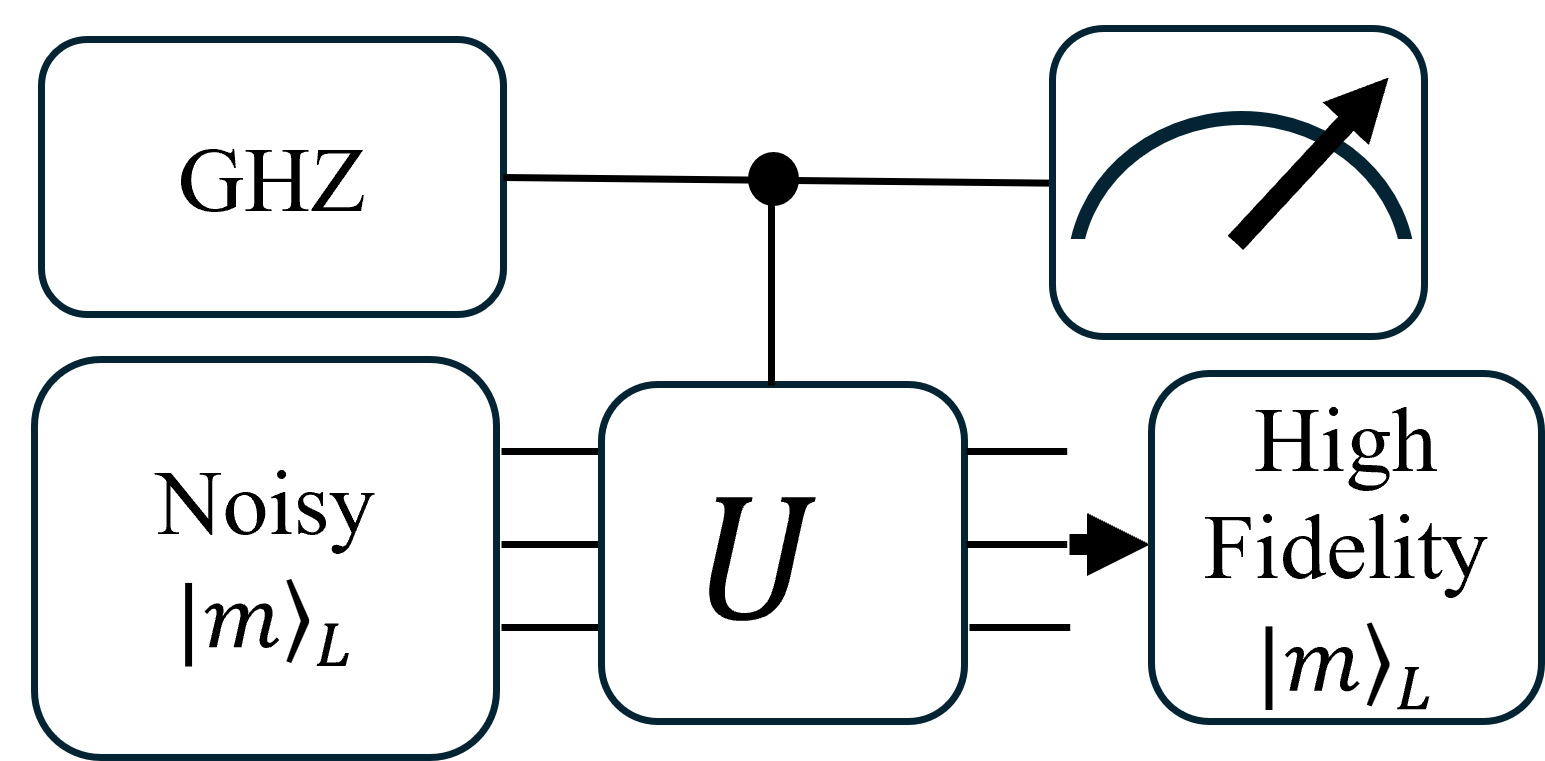}
\caption{\textbf{General phase kickback measurement for arbitrary $\ket{m}_L$.} The $N$ qubit GHZ state $\ket{\psi}=\frac{1}{\sqrt{2}}\left ( \ket{0}^{\bigotimes N}+\ket{1}^{\bigotimes N}  \right )$ is prepared to check the quality of $\ket{m}_L$ by applying tranversal control-$U$ gates with target qubits that are data qubits from the code state and control qubits that are qubits of the GHZ state. For the final measurement, one directly measures all ancilla qubits in the $X$ basis and checks GHZ state parity. If the final measurement outcome on the GHZ state gives odd parity indicating an incorrect logical state $\ket{m}_L$ preparation, one discards the result. Even parity indicates high fidelity logical state $\ket{m}_L$ preparation. Alternatively, one can disentangle the GHZ state back into an all $Z$ product state and measure all qubits in the $Z$ basis. A non-zero result indicates that the state is corrupted by noise and can be discarded.}  
\label{fig:dp}
\end{figure}
The general idea of MSC is to check the quality of magic state preparation $\ket{m}_L$ using phase kickback measurement with a GHZ state~\cite{Chamberland2020,Itogawa2025}. As shown in Fig.~\ref{fig:dp}, $U$ is a transverse gate of the code and also satisfies
\begin{align}\label{eq:u_check}
    U\ket{m}_L = \left ( +1 \right )\ket{m}_L .
\end{align}
Thus, one can filter out non-trivial results based on whether the outcome of the phase kickback measurement gives odd parity or not. The Google team improved this method with a double-phase kickback measurement~\cite{Gidney2024}, which is composed of a phase kickback measurement circuit and its inverse circuit. Furthermore, the magic states are usually prepared as
\begin{align}\label{eq:ms}
    \ket{m}_{L,n} \equiv P_{n}\ket{+}_L.
\end{align}
 where $P_n = \mathbf{diag}\left [ 1,\mathbf{exp}\left ( i\frac{2\pi}{2^n} \right ) \right ]$ allowing one to apply a repeat until success protocol then complete the protocol with logical $S$ gate in the worst case. For instance with $n=3$ magic state preparation, one can choose $U=H_{XY}=\frac{X+Y}{\sqrt{2}}$. The codes that support transversal $H_{XY}$ gates are self-dual codes, such as the $2$D color code \cite{Gidney2024}, $\mathrm{SRP}$-$3$ $(5)$ codes \cite{Chen2025}.

\begin{figure*}[t]
\centering
\includegraphics[width=1\linewidth]{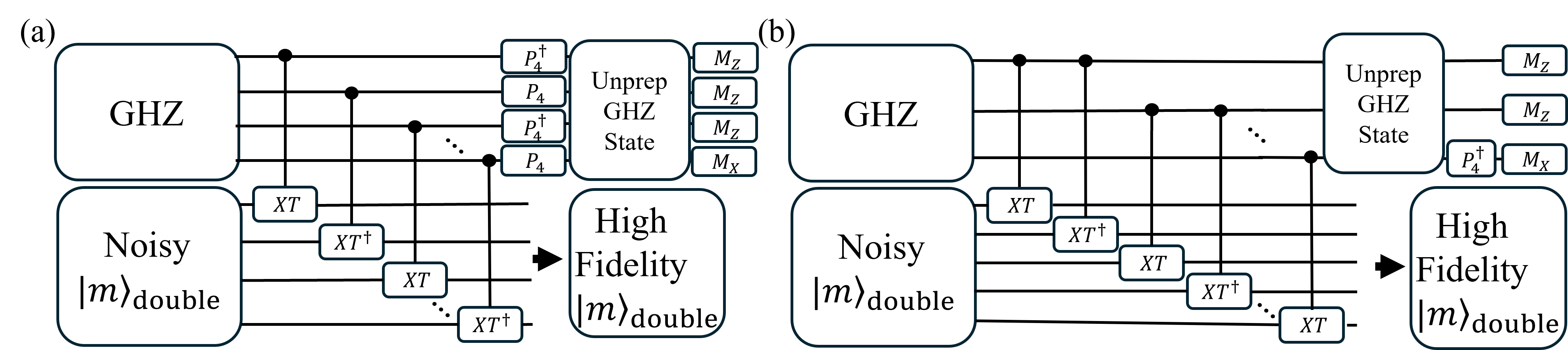}
\caption{\textbf{The phase kickback measurement circuits for $\sqrt{T}\ket{+}_L$ check on doubled color code.} The circuits of phase kickback measurement with (a) phase gates $P_4$ and $P_4^\dagger$ on all GHZ ancilla qubits and (b) a phase gate $P_4^\dagger$ on one ancilla qubit to cancel the phase. After the sequence of control $TX$ and control $T^\dagger X$ gates, the GHZ state is untangled (shown as box with the 'unprep GHZ state' text) into product state with the last qubit in $X$ basis and other qubits in $Z$ basis. If any of measurement outcomes give non-zero results, it indicates the wrong $\sqrt{T}\ket{+}_L$ preparation. Hence, one can directly abandon the shot. At the end, high fidelity magic state are prepared.}
\label{fig:dp_ST}
\end{figure*}

Alternatively, inspired by Ref.~\cite{Sahay2025}, one can perform an $H_{XY}$ check on the surface code, which is composed of a control transversal X gate and a control fold transversal S gate. We extend this method to a more general check for other magic states by replacing the control $S$ gate with the control phase gate
$CP_{n-1}$
so that one can also check for the magic state, $\ket{m}_{L,n} \equiv P_{n}\ket{+}_L$, and cultivate it using 
\begin{align}\label{eq:u_check}
    P_{n-1}X\ket{m}_{L,n} = \mathbf{exp}\left ( i\frac{2\pi}{2^{n}} \right )\ket{m}_{L,n} .
\end{align}
which generates a phase on the ancillas. To cancel the phase, we implement $P_{n}$ phase gates on the GHZ ancilla qubits. (An $n=4$ example is shown in Fig.~\ref{fig:dp_ST}.) Moreover, the code must support a corresponding transversal phase gate $P_{n-1}$ to preserve the code state within the code space during the double-check procedure. To satisfy this condition, the $D$-dimensional color code family is a good candidate for magic state cultivation at the $D$'th Clifford hierarchy level. On the other hand, magic state distillation using a $D$-dimensional color code only produces the magic state at $D-1$'st Clifford hierarchy level~\cite{Kubica2015}. Moreover, MSC is more efficient than magic state distillation.   

\subsection{$\sqrt{T}$ gate Cultivation}

Here, we choose $n=4$ as an example, which corresponds to $\sqrt{T}$ gate cultivation. When $n=4$, the selected code for cultivation must support a transversal $T$ gate. The smallest code that supports a transversal $T$ gate is the $[[10,1,2]]$ code known as Vasmer-Kubica code~\cite{Vasmer2022}. In this case, we need to implement a $4$-qubit $CCCZ$ in order to implement the phase kickback check on [[10,1,2]] code. However, the decomposition of $CCCZ$ into several $2$-qubit gates and single-qubit gates can end up inducing more logical errors for this error detection code. Alternatively, one can use the [[15,1,3]] $3$D code, which supports the transversal $T$ gate by applying physical $T$ and $T^\dagger$ gates, for $\sqrt{T}$ gate cultivation. Similarly, at $d=5$, one can implement a transversal $T$ gate for [[49,1,5]] and [[53,1,5]] codes with physical $T$ and $T^\dagger$ gates. 

The corresponding MSC protocol is as follows:
\begin{enumerate}

\item In the injection stage, we prepare the $\ket{+}_L$ state and apply a noisy $\sqrt{T}$, which can be implemented by two CNOT ladders that sandwich a physical $\sqrt{T}$, in the $3$D code. After the injection, one can apply a round of syndrome measurement to postselect the noisy outcome. 
\item In the cultivation stage, a double-phase kickback check is implemented to filter out incorrect magic states. 
\item We next apply one round of syndrome measurement to postselect noisy outcome. The implementation of phase kickback measurement for $\sqrt{T}\ket{+}_L$ is visualized in Fig.~\ref{fig:dp_ST}. The double-phase kickback check is composed of a phase kickback measurement circuit and the corresponding reversal circuit. If one of measurement results gives $1$, the shot is discarded.
\item During the escape stage, one directly switches to a rotated surface code via lattice surgery (LS) from the $3$D code. Alternatively, one can also code switch to a 2D color code via a one-way transversal CNOT gate~\cite{Daguerre25,Heussen2025}, and transform the $2$D color code to a rotated surface code using LS~\cite{Hirano2025}.
\end{enumerate}Before the escape stage, the high fidelity magic state is prepared on a small doubled color code. To examine the corresponding performance, we simulate $f=3$ and $f=5$ doubled color code cultivation. 

\subsection{f=3}
When $f=3$, the code that supports the transversal $T$ gate is the [[15,1,3]] Quantum Reed-Muller code (More details are in the Appendix~\ref{sec:doubleColorCode}). To initialize $\ket{+}_L$ fault tolerantly, we utilize the MQT QECC package~\cite{mqt,Peham2025} and additional flag qubits for the preparation of the tetrahedral code $\ket{+}_L$~\cite{Forlivesi2025}. Moreover, inspired by Ref.~\cite{Yugo2024}, we also use a Bell pair to extract each syndrome measurement. (The first step gives us the syndrome outcome while the second step provides a flag qubit with information on hook error.) Since there are $10$ $Z$ stabilizers, totally $20$ qubits are required for syndrome measurements. On the other hand, $8$ qubits are required for $4$ $X$ stabilizers.  

As Fig.~\ref{fig:dp_ST} shows, there are two ways to implement the phase kickback measurement for the $\sqrt{T}\ket{+}_L$ check. To cancel the phase in the GHZ ancilla qubits, in Fig.\ref{fig:dp_ST}(a), one applies a phase gate, $P_4^\dagger$ or $P_4$, on each qubit from GHZ state after the control $TX$ or $T^\dagger X$ gates. Thus, in total, a $15$ qubits GHZ state is required. Alternatively, inspired by Ref.~\cite{Sahay2025}, one can also pair up control $TX$ and control $T^\dagger X$ with the same control qubit from the GHZ state, so that the phase accumulated in the GHZ ancilla qubits will be canceled (shown in Fig.~\ref{fig:dp_ST}(b)). One extra remaining control $TX$ gate generates the phase on one of GHZ qubits. Thus, after unpreparing the GHZ state, one still needs to apply a $P_4^\dagger$ gate to cancel the phase. In this case, only a $7$ or $8$ qubit GHZ state is required. Moreover, one can also have the $2$ control $T^\dagger X$ and $2$ control $TX$ operations use the same control qubits as the GHZ state. Hence, a $4$ qubit GHZ state is required in this case.

To estimate the performance of end-to-end $\sqrt{T}\ket{+}_L$ cultivation, we take the same strategy as Ref.~\cite{Gidney2024}: simulate end-to-end $S\ket{+}_L$ cultivation and assume that $\sqrt{T}\ket{+}_L$ cultivation's logical error rate is a given factor of $S\ket{+}_L$'s. However, based on results from Ref.~\cite{Gidney2024}, the performance between $S\ket{+}_L$ and $T\ket{+}_L$ ungrown MSC on the $2$D color code is distinct. In particular, as the noise decreases, the discrepancy between $S\ket{+}_L$ and $T\ket{+}_L$ MSC infidelity becomes larger. Thus, to better understand the difference between $S\ket{+}_L$ and $\sqrt{T}\ket{+}_L$ MSC, we simulate $\sqrt{T}\ket{+}_L$, $T\ket{+}_L$ and $S\ket{+}_L$ ungrown MSC in the $3$D color code under different noise models using state vector simulation. To evaluate the logical infidelity of the magic state $T\ket{+}_L$, $S\ket{+}_L$, and $\sqrt{T}\ket{+}_L$ prepared on $[[15,1,3]]$, we add one extra ideal syndrome measurement after the double phase kickback check to filter out all residual errors. We also apply an ideal logical $T^\dagger$ gate after the ideal syndrome measurement. We then measure the entire doubled color code in the $X$ basis. We calculate the corresponding logical error rate as the number of $1_L$ results divided by the total number of successful shots that yield all-zero syndrome measurements.  

\begin{figure}
   \centering
   \includegraphics[width=1.0\linewidth]{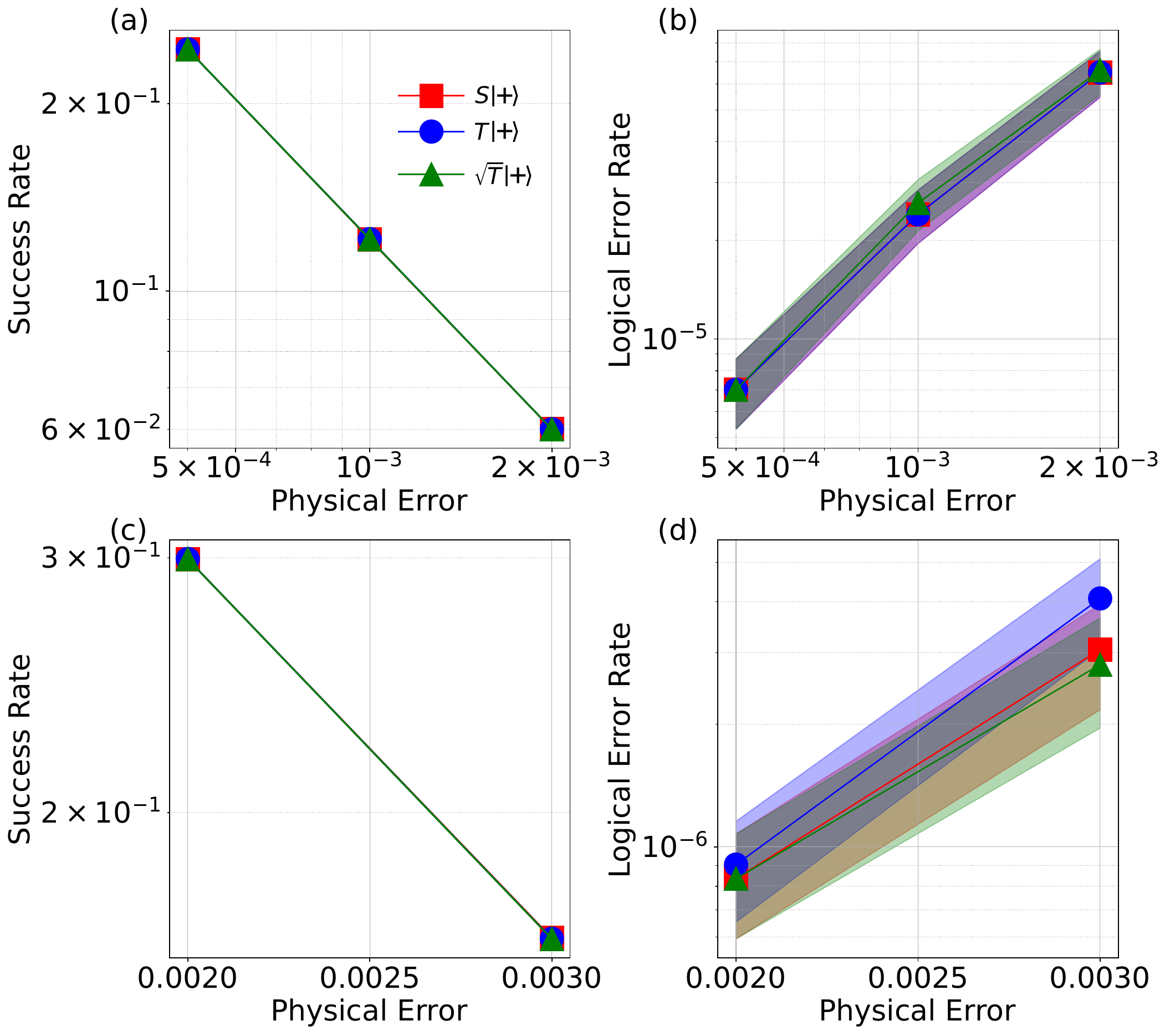}
   \caption{\textbf{The performance of ungrown $d=3$ $\sqrt{T}\ket{+}_L$, $T\ket{+}_L$, and $S\ket{+}_L$ magic state cultivation in statevector simulation.} The success rate (the left panel) and the logical error rate (the right panel) of the ungrown magic state cultivation's statevector simulation. The upper panel is the results with idling noise, while the lower one is the results without the idling noise. The corresponding shades cover the standard error obtained from the sampling.}
   \label{fig:stateVector}
\end{figure}

In Fig.~\ref{fig:stateVector}, the state vector simulations for $\sqrt{T}\ket{+}_L$, $T\ket{+}_L$, and $S\ket{+}_L$ MSCs are shown. With idling noise, the performance of the three states' MSC are highly consistent. On the other hand, without idling noise, the performance of the three states' MSC is a bit different. In particular, for $T\ket{+}_L$, the logical infidelity of the MSCs shows small changes relative to those of the other two MSCs. In contrast, $S\ket{+}_L$ and $\sqrt{T}\ket{+}_L$'s infidelities are more consistent with each other. Note that the variance in logical error rates, here, is due to shot limitations ((b) $8 \times 10^6$ and (d) $5 \times 10^7$). According to the consistent performance between $\sqrt{T}\ket{+}_L$ and $T\ket{+}_L$, we assume that $\sqrt{T}\ket{+}_L$ MSC has similar performance as $S\ket{+}_L$ MSC so that we can use $S\ket{+}_L$ as a proxy state for end-to-end simulation including the code switch and LS using Clifford circuit simulation.         

Before the escape stage, we switch the $[[15,1,3]]$ color code to the $2$D color code so that we can use grafting or LS techniques to convert the color code to the surface code. To understand how best to perform syndrome extraction before and after the phase kickback check in terms of fidelity and efficiency, we simulate the cultivation stage with the following different syndrome measurements and the one way transversal CNOT. Since the code switch between $3$D color code and $2$D color code involves $22$ physical qubits, which is hard to simulate using the state vector simulator, we only simulate the $S\ket{+}_L$ cultivation stage and code switch using Clifford simulation. For infidelity estimation, we use the same method as for state vector simulation. After the rounds of syndrome measurements, we add an extra ideal syndrome measurement and an ideal $S^\dagger$ gate to invert the logical state into $\ket{+}_L$. 

\begin{figure}
    \centering
    \includegraphics[width=1.0\linewidth]{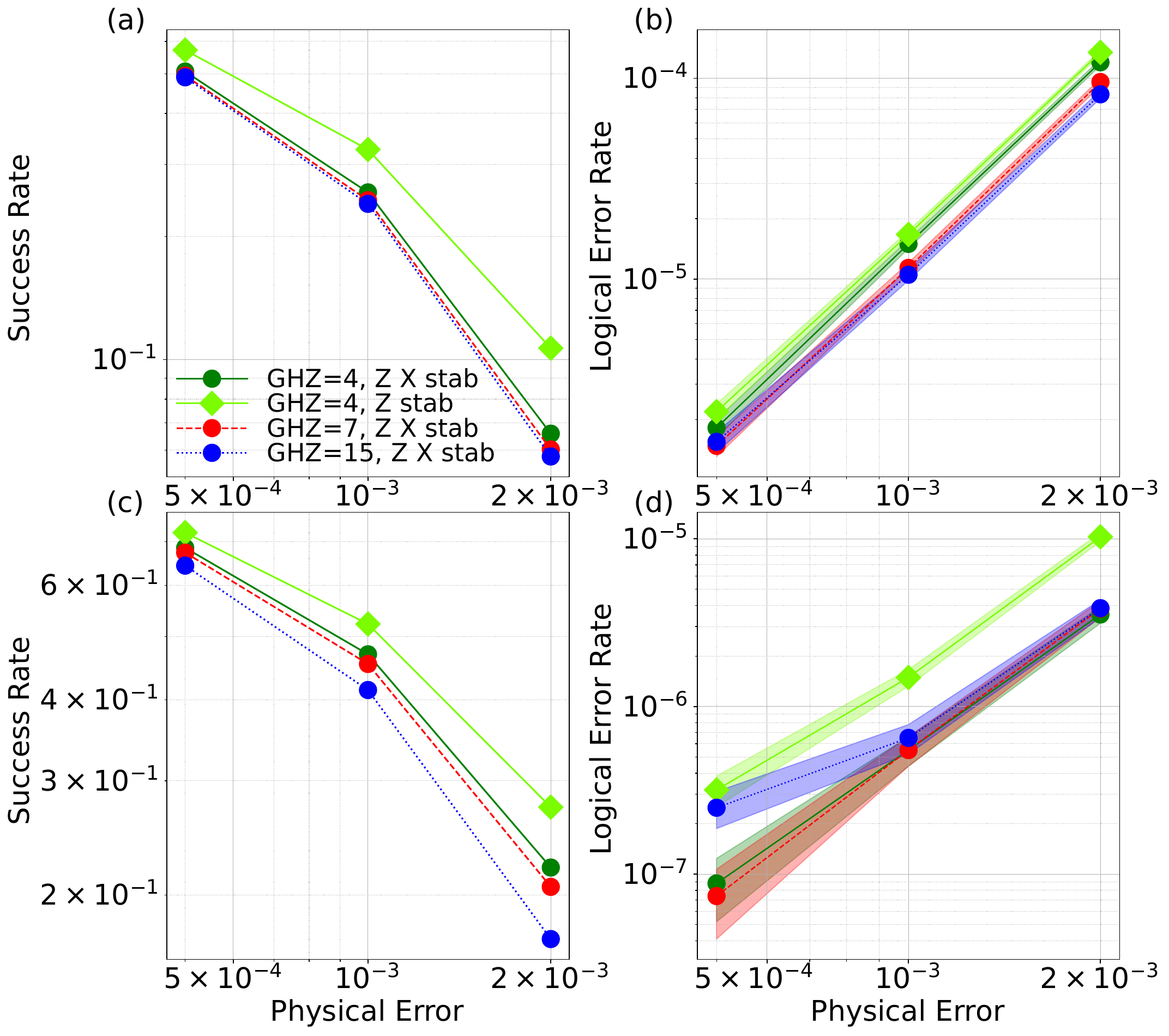}
    \caption{\textbf{Performance of ungrown $f=3$ magic state cultivation using code switching. (With and without idling noise).} The success rate (left panel) and the logical error rate (right panel) varying with physical noise strength. The upper panels (a,b) depict results with idling noise. The lower panels (c,d) depict outcomes without idling noise. In the legend, GHZ indicates how many qubit GHZ states are used. $Z$ stab ($Z$ $X$ stab) represents the process with only $Z$ type (both $Z$ and $X$ type, resp.) stabilizer measurements before and after the double phase kickback check. Corresponding shades represent standard errors obtained from sampling.}
    \label{fig:mainTextUngrown}
\end{figure}

In Fig.~\ref{fig:mainTextUngrown}, we show the performance of ungrown MSC using different GHZ states (15-, 7-, and 4-qubit GHZ states) and different choices of syndrome extraction. With idling noise, using different GHZ states as a double phase check does not yield significantly different logical error outcomes. At a physical error rate of $10^{-3}$, the infidelities of MSC with different GHZ states are around $10^{-5}$. In terms of the success rate, the procedure with only $Z$ stabilizer measurement ($Z$ Stab) before and after the double phase check gives the highest success rate. However, without idling noise, the procedure with only $Z$ stabilizer measurements performs the worst in terms of infidelity, which is above $10^{-6}$ with physical error rate $10^{-3}$. On the other hand, below a physical noise level of $10^{-3}$, both $Z$ and $X$ type stabilizer measurements can generate a magic state with around $7\times10^{-7}$ logical error rate. The results with $4$-qubit and $7$-qubit GHZ states outperform those of $15$-qubit GHZ state in terms of infidelity and success rate. 

\begin{table*}[]
\begin{tabular}{|c|ccc|ccc|}
\hline
\multirow{2}{*}{Tests} & \multicolumn{3}{c|}{49 qubits}                                                                                & \multicolumn{3}{c|}{53 qubits}                                                                                \\ \cline{2-7} 
    & \multicolumn{1}{c|}{Success Rate ($\%$)} & \multicolumn{1}{c|}{Infidelity ($10^{-9}$)} & Std. Dev. ($10^{-9}$) & \multicolumn{1}{c|}{Success Rate ($\%$)} & \multicolumn{1}{c|}{Infidelity ($10^{-9}$)} & Std. Dev. ($10^{-9}$) \\ \hline
w/o $X_{\text{mid}}$        & \multicolumn{1}{c|}{1.903}               & \multicolumn{1}{c|}{6.3}                    & 1.8                  & \multicolumn{1}{c|}{1.483}               & \multicolumn{1}{c|}{2.0}                    & 1.2                  \\ \hline
w/o $X_{\text{aft}}$         & \multicolumn{1}{c|}{1.916}               & \multicolumn{1}{c|}{1.04}                   & 0.74                 & \multicolumn{1}{c|}{1.483}               & \multicolumn{1}{c|}{$\lesssim 0.67$}                  & N/A                \\ \hline
All Stab              & \multicolumn{1}{c|}{1.428}               & \multicolumn{1}{c|}{$\lesssim 0.7$}                  &  N/A               & \multicolumn{1}{c|}{1.080}               & \multicolumn{1}{c|}{0.93}                   & 0.93                 \\ \hline
\end{tabular}
\caption{\textbf{Performance of $f=5$ ungrown magic state cultivation with two double phase kickback checks using $12$ ($13$)-qubit GHZ state.} The performance is estimated under a uniform noise model without idling noise. The procedure involves a code switch from $[[49,1,5]]$ ($[[53,1,5]]$) to the $2$D color code $[[17,1,5]]$ ($[[19,1,5]]$). $X_{\text{mid}}$ indicates $X$ stabilizer measurements in the middle of two double phase kickback checks, while $X_{\text{aft}}$ indicates $X$ stabilizer measurements after the second double phase kickback check. Here, we have a total of $10^{11}$ shots for each case. However, the total shot is not enough to estimate the logical infidelities of some cases. For those cases that does not show any logical errors, we estimate the corresponding logical error rate $\lesssim 1/N_\mathrm{eff}$, where $N_\mathrm{eff}$ represents the total number of successful shots. The corresponding standard errors are not applicable. 
}
\label{tab:mainTextunGrown}
\end{table*}

\begin{figure*}
    \centering
    \includegraphics[width=1.0\linewidth]{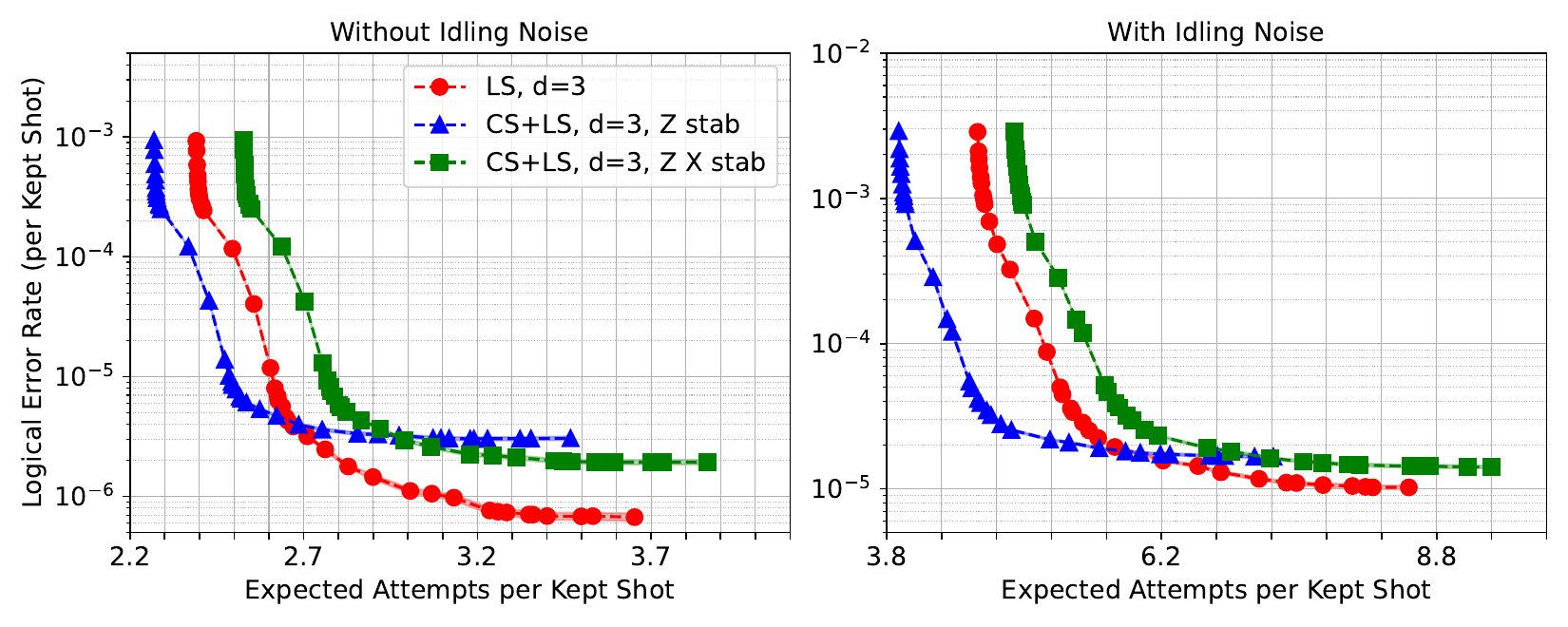}
    \caption{\textbf{End-to-end simulation result for magic state cultivation.} The logical infidelity of simulation for $f=3$ under the uniform noise model $\epsilon = 10^{-3}$ with idling noise (right plot), and without idling noise (left plot). The curves vary with complementary gap cutoff to postselect the logical error-prone sydrome. LS represents the escape stage using direct lattice surgery between rotated surface code ($d_f=11$) and the $[[15,1,3]]$ doubled color code. LS+CS indicates the escape procedure using code switch and lattice surgery between rotated surface code ($d_f=11$) and $[[7,1,3]]$ Steane code.  Z Stab (Z X Stab) represents the process with only Z type (both Z and X type) stabilizer measurements before and after the double phase kickback check. (Standard error is visualized by the shade, which is too narrow to see.)}
    \label{fig:mainTextGrown}
\end{figure*}


\subsection{f=5}
When $f=5$, one can stack $4$ layers of $2$D $d=5$ color code with $1$ qubit on the top into a stacked color code~\cite{Jochym-O'Connor16}. Nevertheless, the corresponding code requires $77$ physical qubits to stack (6,6,6) $d=5$ color codes. Alternatively, with the doubling transformation, one can construct a $d=5$ doubled color code from a $d=5$ $2$D color code~\cite{Bravyi2015,Jain2025}. A [[49,1,5]] code can be obtained from $(4,8,8)$ [[17,1,5]]. Moreover, the [[53,1,5]] code is doubly constructed from $(6,6,6)$ [[19,1,5]]. Here, based on the construction of a recursive, stacked, and capped color code, we construct [[49,1,5]] and [[53,1,5]] codes from $(4,8,8)$ and $(6,6,6)$ $2$D color codes (More details are in Appendix~\ref{sec:dcc}). However, expanding from the [[15,1,3]] color code to the d=5 doubled color code is excessively complicated. We leave the code expansion to the future work. Instead, we directly inject the $d=5$ doubled color code into the noisy target magic state. Nonetheless, the corresponding initial state is very noisy. Thus, we need to perform the double phase kickback check twice to lower the logical error rate. For the code switch, since the $d=5$ doubled color code still share the common $Z$ stabilizers with the $d=5$ $2$D color code, the [[49,1,5]] ([[53,1,5]]) code can also teleport the state's information from doubled color code into the $2$D color code, [[17,1,5]] ([[19,1,5]]), via a one-way transversal $\overline{\text{CNOT}}$ gate~\cite{Matthew2024}. Moreover, one can also consider the $d=5$ tetrahedral $3$D color code $[[65,1,5]]$~\cite{Butt2025}. However, this requires more physical qubits than the doubled color code. So, we only focus on the $d=5$ doubled color code.      

To fault tolerantly initialize $\ket{+}_L$ on the $d=5$ doubled color code, we also use the MQT QECC package~\cite{mqt,Peham2025} and use extra flag qubits to check the quality of $\ket{+}_L$~\cite{Forlivesi2025}. To increase the success rate of $\sqrt{T}$ MSC, similar to the $d=3$ case, we also pair control $TX$ and $T^\dagger X$ to implement the double phase kickback check on the $[[49,1,5]]$ ($[[53,1,5]]$) code, so that we just need to prepare a $25$ ($27$) qubit GHZ state instead of $49$ ($53$) qubit GHZ state. 

Moreover, we can also use $2$ control $TX$ and $2$ control $T^\dagger X$ gates to implement the double phase kickback check, so only a $12$ ($13$) qubit GHZ state is required. In order to maintain the fault distance, we implement the double phase kickback check twice. Before performing an end-to-end simulation, it is essential to test which GHZ state and stabilizer measurement option has the best performance in terms of success rate and logical error rate. Furthermore, we also explore how many syndrome measurements are required before, in between, and after the double phase kickback checks. We employ the same method as the $f=3$ case to evaluate the logical error rate and success rate.

Table~\ref{tab:mainTextunGrown} shows the performance of $f=5$ ungrown cultivation with code switching to the $2$D color code. Without one $X$ stabilizer measurement in the middle of two double phase kicķback checks or after the second double phase kickback check, $53$ qubits MSC outperforms the $49$ qubit result. With all $X$ and $Z$ stabilizer measurements taken before, during, and after the two double phase kickback checks, the MSC performs the best (due to insufficient samples, we just estimate rough infidelities, here), but the corresponding success rate is the lowest. In this case, $49$ qubit MSC performance is slightly better that $53$ qubit MSC. 


\subsection{Growing via Lattice Surgery}
In order to perform quantum error correction on the cultivated code and keep the noise low, it is essential to grow the code into a bigger code, for which the memory logical noise is much lower than the cultivation logical noise. For the $3$D color code, there are no methods proposed to expand into a bigger $3$D color code. However, one can do a code conversion into a rotated surface code using LS~\cite{Chen2026} (more details on LS between doubled color code and surface code are in Appendix~\ref{sec:dcc}). Alternatively, one can switch the $3$D color code into the $2$D color code, and perform LS on the $2$D code with a bigger rotated surface code or graft the $2$D color code onto a large grafting surface code. Finally, one can filter out other corrupted results by calculating the complementary gap from the correlated minimum weight perfect matching (MWPM) decoder~\cite{Gidney2023,Higgott2025}. However, a non-trivial syndrome on the color code cannot be decoded using a correlated MWPM decoder. Hence, we also need to discard the results with the non-trivial syndrome outcome from the color code or the boundary between the color code surface code. Moreover, here, we directly perform LS on the $2$D or doubled color code into a target-sized rotated surface code. So, unlike Ref.~\cite{Hirano2025}, we don't need to do extra expansion on the surface code.


\section{End to End Simulation Result}
To evaluate the performance of the entire MSC process, we simulate the entire circuit from the injection stage to the escape stage, where we perform LS to the target size surface code, outlined above.  However, since $\sqrt{T}$ cultivation involves a non-Clifford gate, increasing the complexity of the simulation, we only simulate the infidelity of $S\ket{+}_L$ MSC and assume that the infidelity of $\sqrt{T}$ MSC is related by an overall multiplicative factor to MSC for  $S\ket{+}_L$'s. 

Fig.~\ref{fig:mainTextGrown} depicts the performance of end-to-end MSC simulation using different escape strategies and different syndrome extraction choices. With idling noise, the best logical infidelity performance that MSC can reach is $10^{-5}$ when using $3$D LS to escape from the small code. Although MSC with $2$D LS and only $Z$ stabilizer measurements has much more success rate to reach $2\times 10^{-5}$, its logical infidelity becomes saturated at around $1.8 \times 10^{-5}$ as the complementary cutoff increases. Without idling noise, $3$D LS can bring MSC infidelity below $10^{-6}$. In contrast, MSC with $2$D LS and both $Z$ and $X$ stabilizer measurements only reaches $2\times10^{-6}$ infidelity. Overall, MSC using $3$D LS as escape strategy performs the best in terms of infidelity.       

In addition to the performance, we also estimate the spacetime volume of the end-to-end simulation. Spacetime volume is defined as the product of the number of active qubits with the circuit depth of each stage of the circuit. Thus, we estimate the expected spacetime volume per retained shot by
\begin{align}\label{eq:stv}
 V=\frac{1}{f_M}\sum^M_{i} f_i V_i
\end{align}
where $f_i$ is the success rate of circuit stage $i$, $V_i$ is the corresponding spacetime volume of that section of the circuit, and $f_M$ is the total success rate of the entire circuit, which includes the complementary gap based postselection. 

\begin{table}[h]
\centering
\begin{tabular}{|l|c|c|c|}
\hline
 & i.d. noise & $V$ & Final logical error rate \\
\hline
 LS & yes & 48236 & $1.5 \times 10^{-5}$ \\
 LS+CS Z X Stab & yes  & 47112 & $1.5 \times 10^{-5}$ \\
 LS+CS Z Stab & yes  & 40009 & $2 \times 10^{-5}$ \\
LS & no & 36477 & $7 \times 10^{-7}$   \\
LS+CS Z X Stab & no & 30185 & $3 \times 10^{-6}$   \\
LS+CS Z Stab & no  & 33870 & $3 \times 10^{-6}$ \\
 \cite{Sahay2025} & yes & $5697$ & $2 \times 10^{-6}$ \\
 \cite{Gidney2024} & yes & $54882$ & $2 \times 10^{-6}$ \\
\hline
\end{tabular}
\caption{\textbf{Expected spacetime volume per successful shot.} The expected spacetime volume per successful shot to achieve a certain logical error rate after complementary gap based postselection for different escape strategies with and without idling (i.d.) noise. LS represents lattice surgery between $d=3$ doubled color code and $d_f=11$ rotated surface code. LS+CS indicates lattice surgery between $d=3$ Steane code and $d_f=11$ rotated surface code. Z X Stab (Z Stab) indicates a cultivation procedure with both Z and X stabilizers (only Z stabilizer, resp.) before and after the double phase check. More details about the expected spacetime volume are in Appendix~\ref{sec: spt_ve}. Magic state cultivation in Ref.~\cite{Sahay2025} is from $d_i=3$ rotated surface code to $d_f=13$ rotated surface code, while magic state cultivation in Ref.~\cite{Gidney2024} is from $d_i=3$ $2$D color code to $d_f=15$ grafting surface code.}
\label{tab:ls_error_rates}
\end{table}

In Table~\ref{tab:ls_error_rates}, the estimated $V_i$ and final logical error rates for different MSC procedures are provided. With idling noise, although the procedure with $2$D LS and both $Z$ and $X$ stabilizer measurements outperforms that with $3$D LS in terms of $V$ to reach infidelity $1.5 \times 10^{-5}$, the procedure with $3$D LS can reach lower infidelity with higher complementary gap cutoff as seen in Fig.~\ref{fig:mainTextGrown}. The procedure with $2$D LS and only $Z$ type syndrome measurement trades off a bit of its infidelity (to $2\times10^{-5}$) to obtain a lower $V$. On the other hand, without idling noise, MSC with $3$D LS can generate $7\times10^{-7}$ magic states with $V=36477$. In contrast, MSCs with $2$D LS only generate logical infidelities above $10^{-6}$. MSC with both $Z$ and $X$ stabilizer measurement has lower $V$ than those with only $Z$ stabilizer measurements, achieving $3\times10^{-6}$. Although the logical error rate of $\sqrt{T}\ket{+}_L$ is higher than current $T\ket{+}_L$ MSC as indicated by the last two rows of the table, high fidelity $\sqrt{T}\ket{+}_L$ still has the potential to shorten the gate sequence. (discussed in sub sec.~\ref{sec:FFTQC})  

\begin{table*}[]
\begin{tabular}{|c|ccc|ccc|}
\hline
\multirow{2}{*}{Tests} & \multicolumn{3}{c|}{49 qubits}                                                                                 & \multicolumn{3}{c|}{53 qubits}                                                                                 \\ \cline{2-7} 
& \multicolumn{1}{c|}{Success Rate ($\%$)} & \multicolumn{1}{c|}{Infidelity ($10^{-9}$)} & Std. Dev. ($10^{-9}$) & \multicolumn{1}{c|}{Success Rate ($\%$)} & \multicolumn{1}{c|}{Infidelity ($10^{-9}$)} & Std. Dev. ($10^{-9}$) \\ \hline
w/o $X_{\text{aft}}$         & \multicolumn{1}{c|}{1.2316}              & \multicolumn{1}{c|}{5.7}                    & 2.1                   & \multicolumn{1}{c|}{0.99308}             & \multicolumn{1}{c|}{4.0}                    & 2.0                   \\ \hline
All true               & \multicolumn{1}{c|}{0.94945}             & \multicolumn{1}{c|}{6.3}                    & 2.6                   & \multicolumn{1}{c|}{0.74312}             & \multicolumn{1}{c|}{9.4}                    & 3.6                   \\ \hline
\end{tabular}
\caption{\textbf{Performance of $d=5$ end-to-end magic state cultivation using $12$ ($13$)-qubit GHZ state.} Performance is estimated under a uniform noise model without idling noise. The procedure involves the code switch from $[[49,1,5]]$ ($[[53,1,5]]$) to $2$D color code $[[17,1,5]]$ ($[[19,1,5]]$). $X_{\text{aft}}$ means $X$ type stabilizer measurements after the second double phase kickback check and the lattice surgery between $d=5$ $2$D color code and $d_f=13$ rotated surface code. Here, we choose the cutoff of the complementary gap around $80$.}
\label{tab: ungrown_d5_e2e}
\end{table*}

In Table~\ref{tab: ungrown_d5_e2e}, the performance of $f=5$ end-to-end simulation under a uniform noise model without idling noise is shown. With fixed complementary gap around $80$, we obtained a logical infidelity of $d=5$ MSC around $5\times10^{-9}$ for different GHZ state choices and stabilizer measurements. The corresponding success rates are around $1$ percent. Compared to $f=3$ MSC, $f=5$ MSC's logical errors are improved by $100$. However, the corresponding success rate is only $1$ percent. To address this issue, based on other MSC protocols, one can perform MSC on $d=3$ first, grow the code to $d=5$, then implement a double phase kickback check on it. We leave this exploration to a future study.      

\section{Application}
In this section, we focus on use cases for $\sqrt{T}$ MSC. Although one can produce $2$ $\sqrt{T}$ gates using the catalyst method with $5$ extra T gates, the magic state catalyst (MSCa) method requires a $\sqrt{T}\ket{+}_L$ as the initial state~\cite{Gidney2019,Kivlichan2020}. Thus, one can implement our method to get $\sqrt{T}\ket{+}_L$ and use it to catalyze more $\sqrt{T}$ gates until the magic state gets too noisy due to memory noise. Alternatively, if the cost, like $V$ in Eq.~\ref{eq:stv}, for $\sqrt{T}\ket{+}_L$ MSC is low, one can just directly use magic state teleportation to implement a $\sqrt{T}$ gate. Although the process still needs $T$ gates to correct some incorrect angle rotations, Clifford+$\sqrt{T}$+$T$ has the potential to shorten the gate synthesis sequence to implement approximate rotation gates~\cite{Kliuchnikov2023}. We discuss the usage of this method for near-term early fault-tolerant and fully fault-tolerant eras.  

\begin{figure}[t]
\centering
\includegraphics[width=1\linewidth]{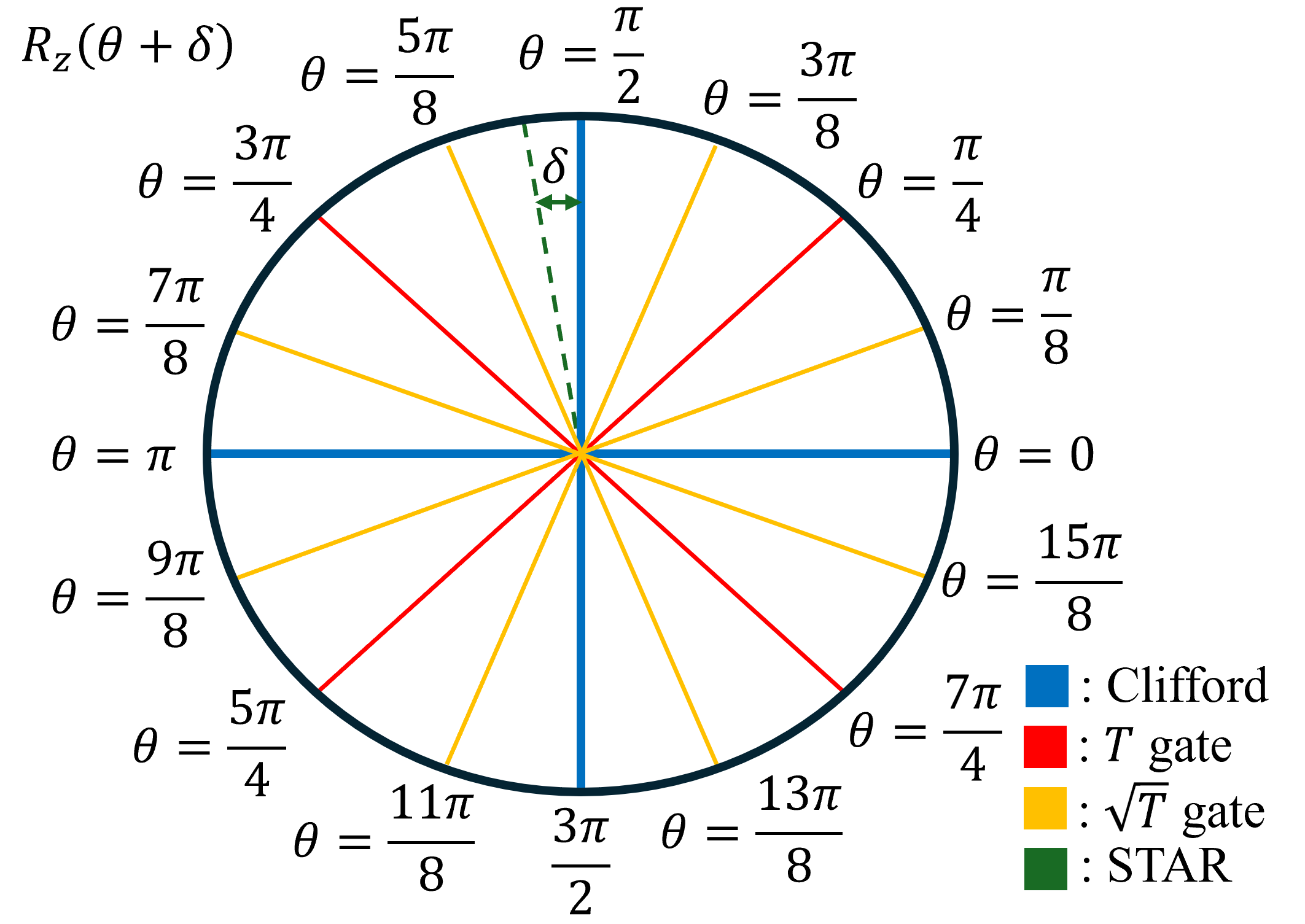}
\caption{\textbf{Illustration of omni-STAR.} The Z rotation angle is decomposed into $\theta$ (large angle, $T$ and $\sqrt{T}$) and $\delta$ (small, injected STAR rotations). $\theta$ is realized with a fault tolerant gate set $ \left\{ Z, S, T, \sqrt{T}\right\}$ where $T$ (red lines) and $\sqrt{T}$ (orange lines) is generated using MSC and the catalyst method. Logical Clifford gates (blue lines) are easily implemented using transversal gate or lattice surgery for the rotated surface code. $\delta$ is realized by the non-fault tolerant STAR rotation (dashed green line).}
\label{fig:O_STAR}
\end{figure}

\begin{figure}[t]
\centering
\includegraphics[width=1.0\linewidth]{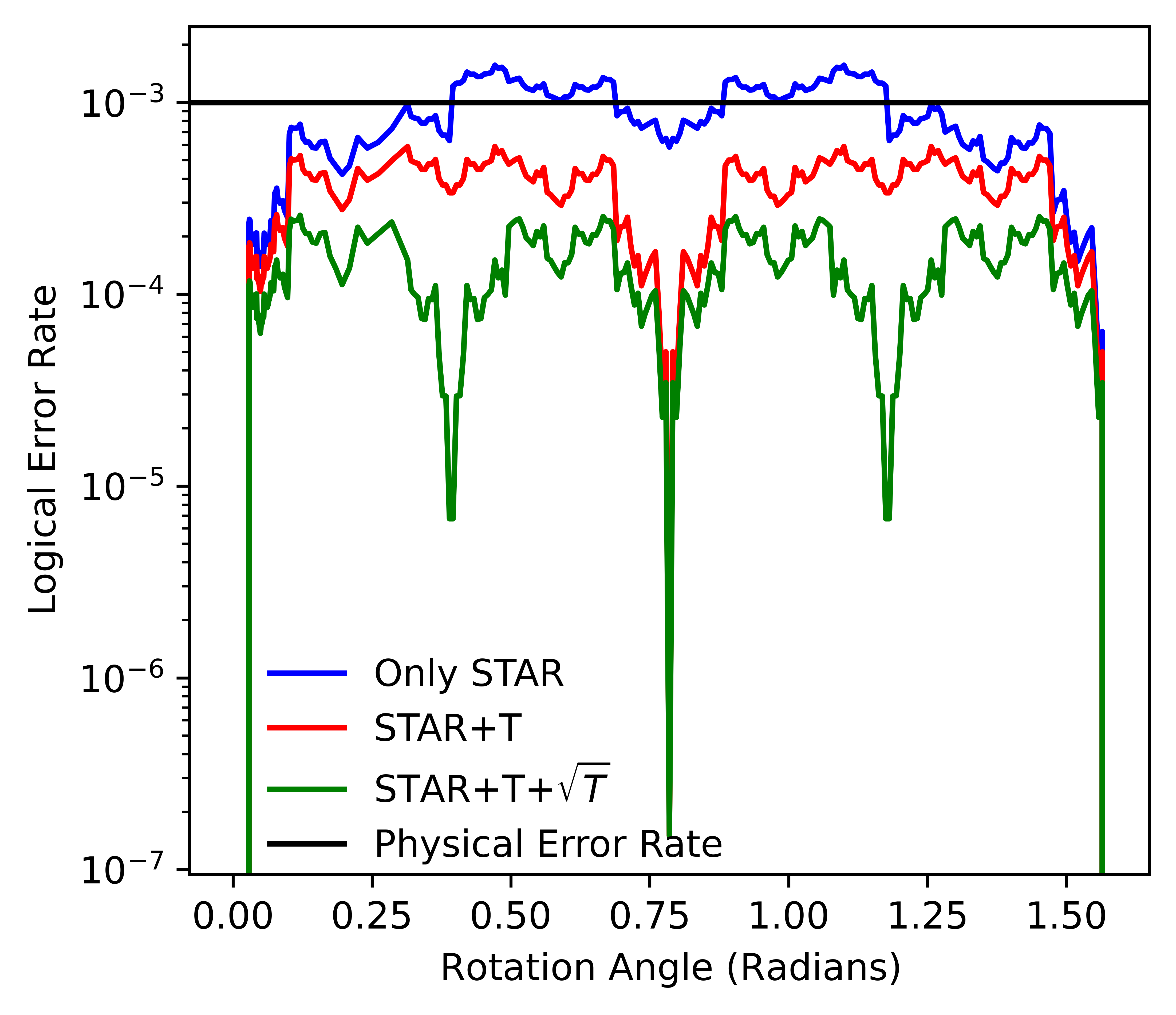}
\caption{\textbf{Logical error rate of arbitrary rotation angle using different gate sets.} The logical error rate for arbitrary angle rotation using only STAR (blue line), STAR+$T$ (red line), and STAR+$T$+$\sqrt{T}$ (green line) is estimated as a function of angle from $0$ to $\pi/2$ radians. The black line represents the physical error rate of a single qubit rotation, which we set to $1\times10^{-3}$ without idling noise. Here, we assume $T$ and $\sqrt{T}$ can be produced via $d=3$ magic state cultivation with corresponding logical error rates of $1.5\times10^{-7}$ and $6\times10^{-7}$.} 
\label{fig:STAR_compr}
\end{figure}
\subsection{Early Fault Tolerant Quantum Computer (EFTQC)}
In the EFTQC era, hardware noise levels are slightly below a QEC code threshold. In this case, it is still too expensive to implement gate synthesis (Clifford+$T$ gate) for rotation gates. Under this scenario, one can implement space-time efficient analog rotation (STAR) for small angle rotation gates~\cite{Toshio2024,ismail2026transversal}. However, if the quantum algorithm requires large angle rotations, the corresponding logical infidelity of STAR rotations is too large due to logical errors proportional to the target rotation angle
\begin{align}
    \epsilon_{L,R}(\theta_r) \approx \alpha p_{\mathrm{ph}}  \left | \theta_r \right | \; ,
\label{eq: el_star}
\end{align}
where $p_{\mathrm{ph}}$ is the physical error rate, $\theta_r$ is the target rotation angle, and $\alpha$ is the constant coefficient. Therefore, we propose \textit{omni-STAR} which involves transversal $Z$ and $S$ gates and $T$ and $\sqrt{T}$ gates produced by MSC factories to address the EFTQC issues mentioned above. In this case, three different kinds of magic state factories are required to implement omni-STAR.

As Fig.~\ref{fig:O_STAR} shows, an arbitrary $R_Z$ rotation angle can be written as $\theta_r=\theta+\delta$ where $\theta$ angle rotation can be implemented by combining fault tolerant gates $ \left\{ Z, S, T, \sqrt{T}\right\}$, where angles $\delta$ satisfying the rotation constraint, $ -\frac{\pi}{16} \leq\delta\leq \frac{\pi}{16}$, can be realized with STAR. Here, the total logical error rate of the rotation angle becomes
\begin{align}
     \epsilon_{L,\text{Tot}}(\theta_r)=\epsilon_{L,F}(\theta) +\epsilon_{L,R} (\delta) \; ,
\end{align}
where $\epsilon_{L,F}$ is the logical noise induced by fault tolerant gates. Based on the Eq.~\ref{eq: el_star}, the maximum logical error rate for the rotation gate is bounded by $\frac{\pi\alpha p_{\mathrm{ph}}}{16}$, where fault tolerant gates' logical error rates are much smaller than this error rate. 

In the original STAR architecture, $\alpha$ is empirically around $1.5$, estimated based on the repeat-until-success (RUS) protocol. In the RUS worst case, $T\ket{+}_L$, $\sqrt{T}\ket{+}_L$ are required to cancel incorrect rotation angles and are noisy if one produces those states using STAR. On the other hand, for omni-STAR, one uses MSC to generate $T\ket{+}_L$, $\sqrt{T}\ket{+}_L$ magic states. Thus, the logical error rate for omni-STAR with RUS becomes  
\begin{align}
     \epsilon_{L,R}&\left ( \delta=\frac{\pi}{2^{n}} \right )=\frac{1}{2^{n-1}}\epsilon_{L,S}+\frac{1}{2^{n-2}}\epsilon_{L,T}+\frac{1}{2^{n- 3}}\epsilon_{L,\sqrt{T}} \notag\\
     &+\sum_{m=0}^{n-2}\frac{m}{2^m}\epsilon_{L,\text{CNOT}}+\sum_{m=0}^{n-4} \frac{1}{2^m} \epsilon_r\left ( \theta=\frac{\pi}{2^{n-m}} \right )
\end{align}
where $\epsilon_{L,G}$ is the logical error for implementing $G$ gates with $G\in \left\{ S, T, \sqrt{T},\text{CNOT}\right\}$, and $\epsilon_r\left ( \theta \right )$ is the logical error rate of STAR injection and teleportation for $R_z(\theta)\ket{+}_L$. For other arbitrary rotation angles, $\delta$ can be a linear combination of $\frac{\pi}{2^k}$ discrete angles $\delta=\sum_k C_{k,\delta}\frac{\pi}{2^k}$ with $C_{k,\delta}$ the integer coefficients for linear combination satisfying $C_{k,\delta} \in \left\{ 1, 0, -1\right\}$. Thus, the total logical error rate for omni-STAR for arbitrary angle is
\begin{align}
     \epsilon_{L,R}\left ( \delta \right )=\sum_k C_{k,\delta} \epsilon_{L,R}\left ( \frac{\pi}{2^k} \right )
\end{align}
With the formula above, we can estimate the logical error rate of arbitrary angle rotations using omni-STAR and STAR-only. 

Here, we implement STAR and omni-STAR on a $d=11$ rotated surface code and apply partial error correction after $k=3$ multi-transversal rotation (more details in Appendix). Moreover, to optimize the performance of STAR, we also implement the switch protocol mentioned in Ref.~\cite{Toshio2024}. When $k=3$, multi-transversal rotation's logical error rate is higher than $k=1$'s, so we instead implement $k=1$ multi-transversal rotation for magic state preparation. We also consider the logical error rate of logical CNOT gates using LS. Based on the result from Ref.~\cite{Gidney2024}, when $d=11$, the logical error rate of an LS based CNOT on the surface code is roughly $6 \times 10^{-7}$. The logical CNOT is used for the teleportation of the magic state. 

As Fig.~\ref{fig:STAR_compr} shows, when the rotation angle is around $0.3-0.7$ or $0.8-1.2$ radians, the performance of STAR is worse than with physical rotational angle gates. With fault tolerant gates generated from an MSC factory, logical error rates for any arbitrary rotation angle are below physical rotation error rates. Generally, omni-STAR logical error rates are around $5 \times 10^{-4}$. When the rotation angle is near $\pi/4$, $\pi/2$ or $0$, the logical error rate can reach below $10^{-4}$. Furthermore, with $T$ and $\sqrt{T}$ gates, the logical error rate becomes around $1 \times 10^{-4}-2 \times 10^{-4}$. On average, compared to only using STAR, the logical error rate for omni-STAR (STAR+$T$+$\sqrt{T}$) is around $2\times-10\times$ improved. 

For further improvement, one can use a $T^{1/4}$ catalyst and use $T$, $\sqrt{T}$ gates to obtain $T^{1/4}\ket{+}_L$ as the initial state of the catalyst (replace $T$ gate with $\sqrt{T}$ in fig.~\ref{fig:catalyst}). During the process of $T^{1/4}$ catalysis, $4$ T gates and a $\sqrt{T}$ gate are required. However, to generate a $T^{1/4}\ket{+}_L$, it requires a long sequence of gates, including $T$ gate and $\sqrt{T}$ gate. Instead, one can develop $T^{1/4}$ MSC. In this case, how to implement the injection and escape stages remains an open question that we leave for future work.  

\subsection{Fully Fault Tolerant Quantum Computer (FFTQC)}
\label{sec:FFTQC}
In the FFTQC era, the hardware noise level will be much lower than the threshold. This will allow the implementation of gate synthesis to produce high-fidelity operations using a universal and fault-tolerant discrete set of quantum gates. Clifford + $T$ is the most studied set of universal gates in the literature. However, other possibilities are also the subject of discussion in the fault-tolerant quantum computing community.  

An efficient MSC of  $\sqrt{T}$ along with the production of high-fidelity $T$ gates (\textit{e.g.}, using MSC of $T\ket{+}_L$) will allow us to implement faster universal quantum computation with Clifford + $T$ + $\sqrt{T}$~\cite{Kliuchnikov2023}. In particular, $\sqrt{T}$ gates can be efficiently implemented using magic state teleportation (MST) and/or magic state catalysis (MSCa). Examples of the usage of $\sqrt{T}\ket{+}_L$ include the Fast Fermionic Fourier Transform (FFFT) which requires $\sqrt{T}$ and $\sqrt{T}^3$  gates for some specific applications, as presented by \cite{Kivlichan2020}, and in a more general picture both the FFFT  and even the regular Quantum Fourier Transform requires other single-qubit diagonal rotations (\textit{i.e.} $z$-rotation: $e^{-i\frac{\theta}{2} Z}$ for different $\theta$s) which could be synthesized using Clifford + $ T$+$\sqrt{T}$, as discussed in \cite{Kliuchnikov2023}.  However, it is usually considered the application of $\sqrt{T}$ gates by using MSCa, which highlights two points of discussion: 

\begin{enumerate}
    \item The production of a seed for the catalysis.

    \item The cost of the catalysis method which, to the best of our knowledge, demands 2.5 $T$ gates for each $\sqrt{T}$ gate (Appendix A3 of \cite{Kivlichan2020}).
\end{enumerate}
For instance, seed production is a direct application for MSC of $\sqrt{T}\ket{+}_L$. 

Regarding the second point, if the cost for MSC of a $\sqrt{T}\ket{+}_L$ state is sufficiently small compared to the cost of preparing the $T\ket{+}$ state, which is equivalent to the cost to apply a $T$ gate when counting for non-Clifford resources, then MST could be used as an alternative to MSCa since it requires just one $\sqrt{T}\ket{+}_L$ and one $T$ gate. 

We will address this question from two perspectives. First: a single application of a $\sqrt{T}$ or $\sqrt{T}^3$. Second: the unitary synthesis of a $z$-rotation. 

We are also considering the gate $\sqrt{T}^3$ because it appears in one specific case of the FFFT and also in the unitary synthesis of $z$-rotations using Clifford + $T$ + $\sqrt{T}$ universal set.

\begin{figure}[t]
\centering
\includegraphics[width=1\linewidth]{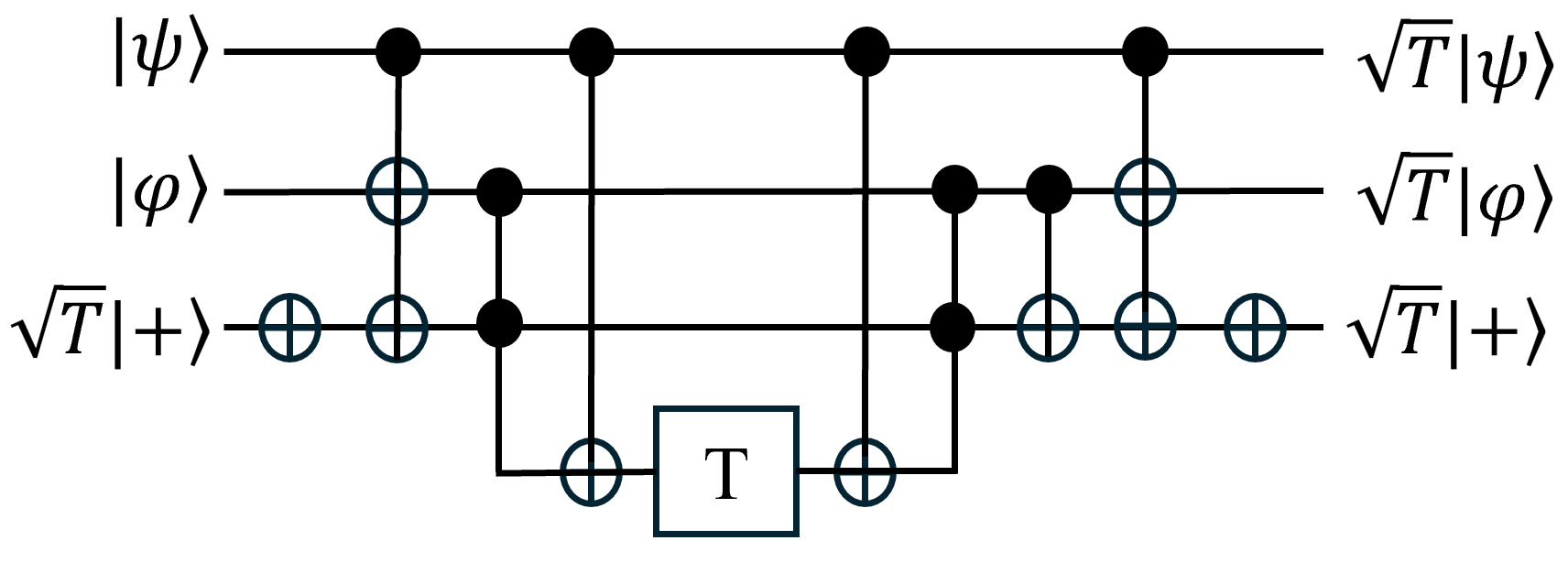}
\caption{\textbf{Circuit for the $\sqrt{T}$ gate catalyst.} The catalysis ``consumes'' a $T$ gate (middle bottom of the circuit) to apply the $\sqrt{T}$ gate to two arbitrary states. A seed state $\sqrt{T}\ket{+}_L$ is necessary to perform the catalysis. In the end of this process the seed state is not consumed allowing it to be reused in posterior catalysis. Besides the consumed $T$ gate, the circuit requires computing and uncomputing the logical AND operation. The former can be decomposed in the Clifford+T universal gate set by using 4 $T$ gates (Figure \ref{fig:catalyst_2}), so the total non-Clifford cost for the catalyst method is 5 $T$ gates. As the seed only needs to be prepared once, we disregard its cost.}
\label{fig:catalyst}
\end{figure}

Considering just the non-Clifford cost, MST uses one $\sqrt{T}\ket{+}_L$ state and demands one $T\ket{+}$ state half of the time to produces one $\sqrt{T}$ gate. So each $\sqrt{T}$ gate costs on average the cost for preparing the $\sqrt{T}\ket{+}_L$ state plus half the cost to prepare a $T$ gate. A circuit for MST of $\sqrt{T}$ is presented in Figure \ref{fig:mst}. 
On the other hand, MSCa demands one $\sqrt{T}\ket{+}_L$ and five $T$ gates to apply two $T$ gates, however it does not consume the seed state $\sqrt{T}\ket{+}_L$ which can be used again in posterior catalysis. The circuit for MSCa is presented in Figure \ref{fig:catalyst} and a detailed decomposition of the circuits which compute and uncompute logical AND operations, necessary for the catalyses, are presented  in Figure \ref{fig:catalyst_2}.

Here, we define the cost for preparing a $\sqrt{T}\ket{+}_L \left(T\ket{+}_L\right)$  as the expected spacetime volume per retained shot, $V_{\sqrt{T}} \left(V_{T}\right)$, of the MSC protocol. The total non-Clifford gate cost for the implementation of $K$ $\sqrt{T}$ gates is, on average:
\begin{align*}
\begin{split}
    \text{MST cost: }  &\quad KV_{\sqrt{T}}+\frac{K}{2}V_{T} \\
    \text{MSCa cost: } &\quad V_{\sqrt{T}}+\frac{5K}{2}V_{T} \; .
\end{split}
\end{align*}
And with some simple algebra and disregarding the factors that are constant with respect to $K$, MST is advantageous over MSCa when 
\begin{align}
    V_{\sqrt{T}} < 2V_{T} \; .
    \label{eq:advantageCondOne}
\end{align}
With respect to $\sqrt{T}^3$, we can implement $\sqrt{T}^3 \ket{+}_L$ cultivation instead by injecting the $\sqrt{T}^3 \ket{+}_L$ and replacing the control $TX$ gate with a control $T^3X$ gate on the double phase kickback check. So, for MST, we can directly teleport the $\sqrt{T}^3$ gate. If the teleportation ($50\%$ chance) fails, we can implement $T^3$ MSC to apply a $T^3$ gate. If the teleportation fails again, we just need to implement the logical Clifford $S^\dagger$ gate. MSCa can be adapted to produce $\sqrt{T}^3$ simply by replacing the seed state $\sqrt{T}\ket{+}_L$ with $\sqrt{T}^3\ket{+}_L$ and adding one Clifford $S$ gate after the $T$ gate in Figure \ref{fig:catalyst}, as discussed in \cite{Kivlichan2020}. Here, we assume the cost of $\sqrt{T}^3\ket{+}_L$ ($T^3\ket{+}_L$) MSC is similar to that of $\sqrt{T}\ket{+}_L$ ($T\ket{+}_L$) MSC. Under these assumptions, the condition for MST be advantageous over MSCa for preparing a $\sqrt{T}^3$ gate is the same as for $\sqrt{T}$, \textit{i. e.} condition \ref{eq:advantageCondOne}. As showed by Table \ref{tab:ls_error_rates}, the logical error rate and spacetime volume of MSC of $\sqrt{T}$ are both worse by about a factor $10\times$ than MSC of $T$ yet, indicating that further improvements in MSC of $\sqrt{T}$ are necessary in order to turn MST a viable option to implement $\sqrt{T}$ gates.

Table \ref{tab:cost} presents the spent resources, the gates and states produced by both methods, considering $\sqrt{T}$ and $\sqrt{T}^3$ and disregarding the cost for seed production for MSCa because it is only prepared once until the seed state becomes noisy.

\begin{table}[]
\begin{tabular}{|cc||cc||c|}
\hline
\multicolumn{2}{|c||}{Method}       & \multicolumn{2}{c||}{Resources}                     & Product      \\ \hline
\multicolumn{1}{|c|}{\multirow{2}{*}{$\sqrt{T}$}} & MST  & \multicolumn{1}{c|}{$\frac{1}{2}T$} & $\sqrt{T}\ket{+}_L$                       & $\sqrt{T}$            \\ \cline{2-5} 
\multicolumn{1}{|c|}{}                       & MSCa & \multicolumn{1}{c|}{5T}   & $0$                       &  $2\sqrt{T}$            \\ \hline
\multicolumn{1}{|c|}{\multirow{2}{*}{$\sqrt{T}^3$}} & MST  & \multicolumn{1}{c|}{$\frac{1}{2}T^3$} & $\sqrt{T}^3\ket{+}_L$                       & $\sqrt{T}^3$             \\ \cline{2-5} 
\multicolumn{1}{|c|}{}                       & MSCa & \multicolumn{1}{c|}{5T}   & $0$                       & $2\sqrt{T}^3$             \\ \hline
\end{tabular}
\caption{\textbf{Average cost for preparing $\sqrt{T}^k, k=1,3$ gates using MST and MSCa.} MSCa demands more $T$ gates than MST, however, it also produces two times more $\sqrt{T}^k$ gates and does not consume the seed (which just needs to be prepared once), making it advantageous when an efficient protocol to produce high-fidelity $\sqrt{T}^k\ket{+}_L$ magic states is not available, whereas MST is the best option if it is possible to produce $\sqrt{T}^k\ket{+}_L$ at low cost. }
\label{tab:cost}
\end{table}

Regarding the universal gate set Clifford +$T$ + $\sqrt{T}$ in comparison with Clifford + $T$ for single qubit rotation synthesis, Ref.~\cite{Kliuchnikov2023} considers two scenarios: Random angles between $0$ and $2\pi$ and Fourier angles $\frac{\pi}{2^k}$ with $k$ integers. They considered four approaches for synthesis and the best one for both (``mixed fallback") obtained a scaling for the non-Clifford gate count for random angles, given a precision $\varepsilon$, of $0.23\log_2\left(1/\varepsilon\right) + 2.13$ for the Clifford + $\sqrt{T}$ and of $0.53\log_2\left(1/\varepsilon\right) + 4.86$ for Clifford + $T$. The scaling for the Fourier angles is roughly similar for high accuracy cases; however, for low accuracy cases, the results from the method approximate the rotation with an identity, $I$, in many instances, generating a large variance in the scaling results.

While these methods make the synthesis using only $T$ gates for the Clifford + $T$, the decomposition is made using $\sqrt{T},T$ and $\sqrt{T}^3$ for Clifford + $T$ + $\sqrt{T}$, so a fair comparison should consider the cost to produce each one of these three gates. If the three types of gate had the same cost, the best metric would be the non-Clifford gate count and the reduction presented in the last paragraph would be achieved. However, when this is not true, a more robust analysis is necessary. The paper takes this into account using the $T$-count for the Clifford + $\sqrt{T}$ considering that $\sqrt{T}$ and $\sqrt{T}^3$ can be produced using catalysis and disregarding the cost for seed preparation. With these assumptions, only the number of $T$ gates scales with the problem since the seeds just need to be prepared once. The catalysis protocol uses four $T$ gates for each $\sqrt{T}$ and $\sqrt{T}^3$ gates, which raises the cost for  Clifford +$T$ + $\sqrt{T}$ to be slightly larger than Clifford + $T$. Using a better protocol, such as the one from Appendix A3 of \cite{Kivlichan2020}, it is possible to optimize this $T$-count. More than that, if the cost for MSC of $\sqrt{T}\ket{+}_L$ satisfies conditions~\ref{eq:advantageCondOne} the cost for the synthesis using Clifford + $\sqrt{T}$ could be reduced even further, getting closer to the one presented in the non-Clifford gate count.

\section{Conclusion}
In this paper, we generalize MSC methods for the different Clifford hierarchy gates by extending the concept from Ref.~\cite{Sahay2025}. We focus on $\sqrt{T}\ket{+}_L$ MSC as a key example. To cultivate the $\sqrt{T}$ gate, the small code that one chooses must support the transversal $T$ gate so that one can construct the corresponding double phase kickback check. We select $[[15,1,3]]$ for $f=3$ and $[[49,1,5]]$, $[[53,1,5]]$ for $f=5$. We also provide two escape strategies for these codes: (1) a dimension jump from $3$D to $2$D, and then LS from the $2$D color code to the rotated surface code, and (2) direct LS between the $3$D color code and a target size rotated surface code. We also test consistency between $S\ket{+}_L$, $T\ket{+}_L$, and $\sqrt{T}\ket{+}_L$ to validate our Clifford simulation. Our end-to-end simulation shows $d=3$ MSC can generate $10^{-5}$ infidelity states under idling noise and $6\times10^{-7}$ without idling noise. For $d=5$, without idling noise, MSC can generate states with infidelity $2.6\times10^{-9}$. The corresponding expected spacetime volume per retained shot for $d=3$ is comparable with the original MSC \cite{Gidney2024}.       

Besides the demonstration of $\sqrt{T}\ket{+}_L$ MSC, we also discuss the use cases of early and fully fault-tolerant quantum computers. In the near term, $T$, $\sqrt{T}$ and Clifford gates, plus the recently proposed small angle rotation technique (STAR), can help to implement certain rotation angle gates with low noise. When physical gate errors are sufficiently low, one can still implement the $T$, $\sqrt{T}$, and Clifford gates to synthesize the rotation $Z$ gate with a potentially shorter sequence of gates, compared to the Clifford+$T$ gate set. We consider two schemes: (1) When the spacetime volume of $\sqrt{T}\ket{+}_L$ MSC is much worse than $T\ket{+}_L$ MSC, one can use MSCa, which needs resource state $\sqrt{T}\ket{+}_L$ to implement $\sqrt{T}$. (2) When the spacetime volume of $\sqrt{T}\ket{+}_L$ MSC is slightly worse than $T\ket{+}_L$ MSC and one can use MST to implement $\sqrt{T}$. 

This work provides MSC schemes beyond $T$ MSC. In particular, the $\sqrt{T}$ MSC method we propose here is also near-term and hardware-friendly, especially for neutral-atom and trapped-ion systems. Here, we grow the small QEC code into the larger size rotated surface code. Alternatively, one can implement unfolding distillation methods to transform $\sqrt{T}\ket{+}_L$ $[[15,1,3]]$ code to the rectangular surface code or repetition code if the physical qubits suffer more from bias error~\cite{Ruiz2025}. Moreover, the final target code does not necessarily have to be the surface code. One can build up the adapter between the color code and quantum low density parity check (QLDPC) code to transform information from the color code to the QLDPC code~\cite{Xu2025}.        

Although omni-STAR can reach arbitrary angles at low cost and with little logical noise, other methods for specific-angle rotations can optimize them, but they are too noisy when using omni-STAR. One solution is to implement Clifford$+T+\sqrt{T}$ synthesis to achieve larger angle rotation~\cite{toshio2026}, for which the STAR injection method has poor performance. Moreover, one can explore the $T^{\frac{1}{4}}$ cultivation using a $4$D color code, enabling omni-STAR to reach arbitrary angles with lower noise. Once the $4$D color code has the same $Z$ stabilizer as the $2$D color code, one can also switch the code and the LS to the rotated surface code. Moreover, $T^{\frac{1}{4}}$ potentially provides a shorter sequence for gate synthesis, which could also be useful for the fully fault-tolerant era. Finally, one can extend this method to $\ket{m,n+1}$ MSC on the code supporting the transversal $P_n$ gate. For the escape stage, one can employ a dimension jump from the high dimension code to the low dimension code of which the escape strategy has been investigated, if one cultivates the magic state on the high dimension color code. Although the injection stage remains unknown, these open questions provide a fruitful playground for future work.

\begin{figure}[t]
\centering
\includegraphics[width=1\linewidth]{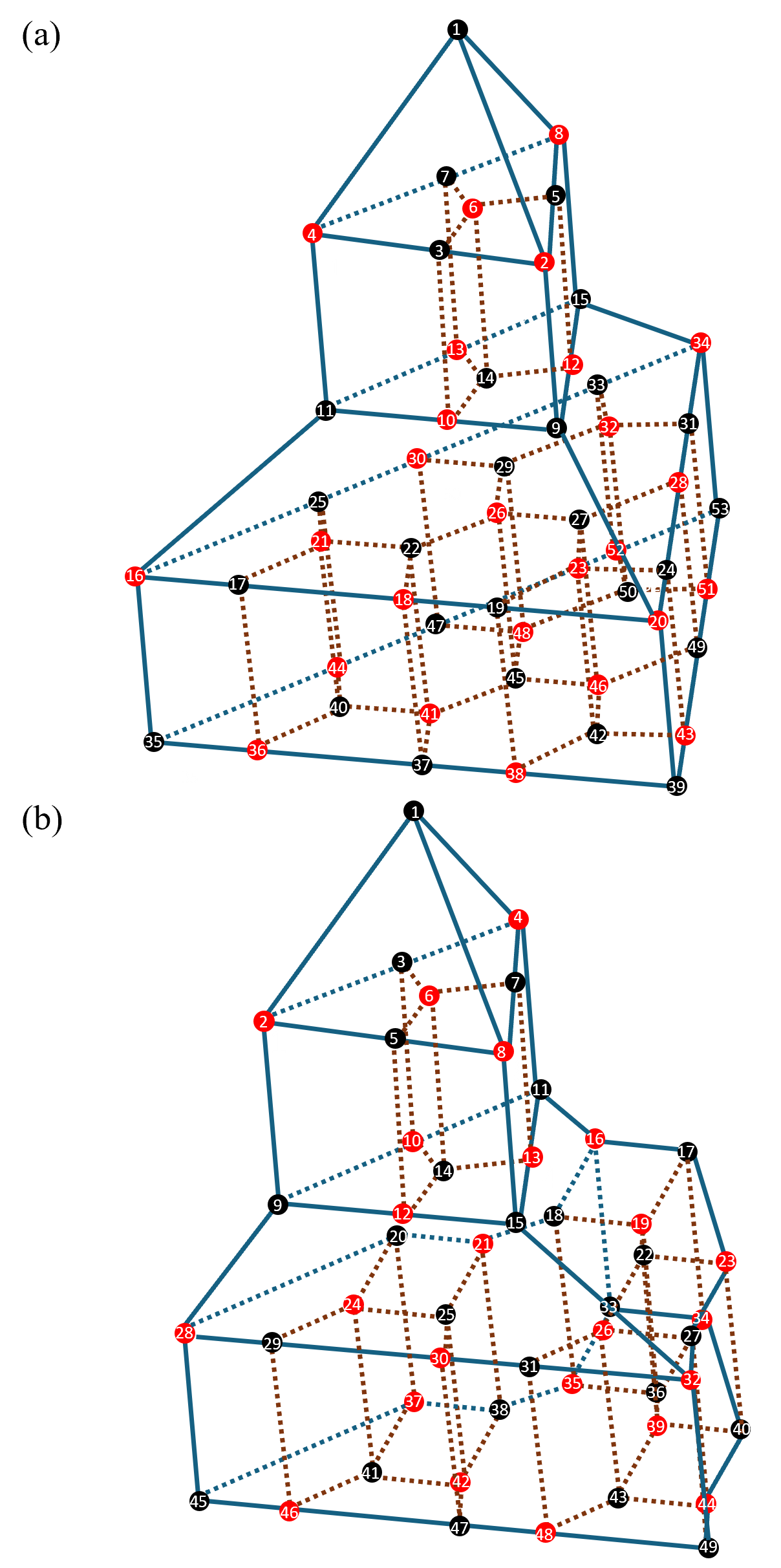}
\caption{\textbf{Visualization of the recursive capped color code.} (a) [[49,1,5]] code constructed from [[17,1,5]] code. (b) [[53,1,5]] code constructed from [[19,1,5]] code. To implement the transversal $T$ gate on those two codes, one can apply physical $T$ gates on the black qubits and $T^\dagger$s on the red qubits.}
\label{fig:d_5_3D_code}
\end{figure}
\begin{acknowledgments}
We acknowledge valuable discussions with Chen Zhao, Pei-Kai Tsai, Kaavya Sahay, Milan Kornjača, Diego García-Martín, and Ivan Chernyshev. This work was supported by the NNSA ASC Beyond Moore's Law project (I.C.C. and A.T.S.). M.S.F. was supported by the Laboratory Directed Research and Development (LDRD) program of Los Alamos National Laboratory (LANL) under project number 20260043DR and São Paulo Research Foundation under process numbers 2023/18240-0 and 2025/11307-7. The LANL designation for this manuscript is LA-UR-26-24630.
\end{acknowledgments}

\appendix

\section{Kickback Phase Measurement}

\begin{figure}[t]
\centering
\includegraphics[width=0.8\linewidth]{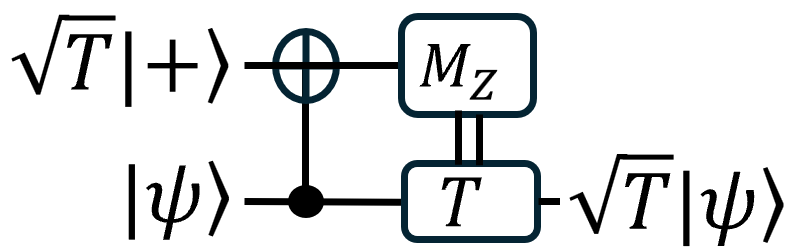}
\caption{\textbf{The circuit for the $\sqrt{T}$ gate teleportation.} Using a magical state $\sqrt{T}\ket{+}_L$ state and a $T$ gate conditioned to the $+1$ result in a $Z$ measurement it is possible to teleport the action of the $\sqrt{T}$ from the $\ket{+}_L$  state to an arbitrary state $\ket{\psi}$. This process consumes the magical state.}  
\label{fig:mst}
\end{figure}

Here, we provide further details on the implementation of the phase-kickback check for $\sqrt{T}\ket{+}_L$, as shown in Fig.~\ref{fig:dp_ST}. An $n$-qubit GHZ state and the logical state $\sqrt{T}\ket{+}_L$ encoded in a specific QEC code are prepared as
\begin{align}
     \ket{\text{GHZ}}_n\otimes \sqrt{T}\ket{+}_L
     = \frac{1}{\sqrt{2}}\left[\ket{0}^{\otimes n}+\ket{1}^{\otimes n}\right]\otimes \sqrt{T}\ket{+}_L .
\end{align}
After the transversal controlled-$XT$ gates are applied, the corresponding state becomes
\begin{align}
    \frac{1}{\sqrt{2}}\left[\ket{0}^{\otimes n}+e^{i\frac{\pi}{8}}\ket{1}^{\otimes n}\right]\otimes \sqrt{T}\ket{+}_L .
\end{align}


To cancel the phase $e^{i\frac{\pi}{8}}$, one can apply $\sqrt{T}$ and $\sqrt{T}^\dagger$, which correspond to $P_4$ and $P_4^\dagger$, respectively, as shown in Fig.~\ref{fig:dp_ST}(a). Alternatively, one can pair controlled-$XT$ and controlled-$XT^\dagger$ gates, as shown in Fig.~\ref{fig:dp_ST}(b), to cancel the phase. Finally, the GHZ state can be disentangled into product states in the $Z$ basis and then measured. Similarly, if the QEC code supports a transversal $P_n$ gate, one can prepare a noisy $P_{n+1}\ket{+}$ state and use controlled transversal-$XP_n$ gates to perform either the phase-kickback check or the double phase-kickback check. In this way, magic states at higher levels of the Clifford hierarchy can be prepared.

\section{Doubled Color Code}
\label{sec:doubleColorCode}

\begin{figure}[t]
\centering
\includegraphics[width=0.7\linewidth]{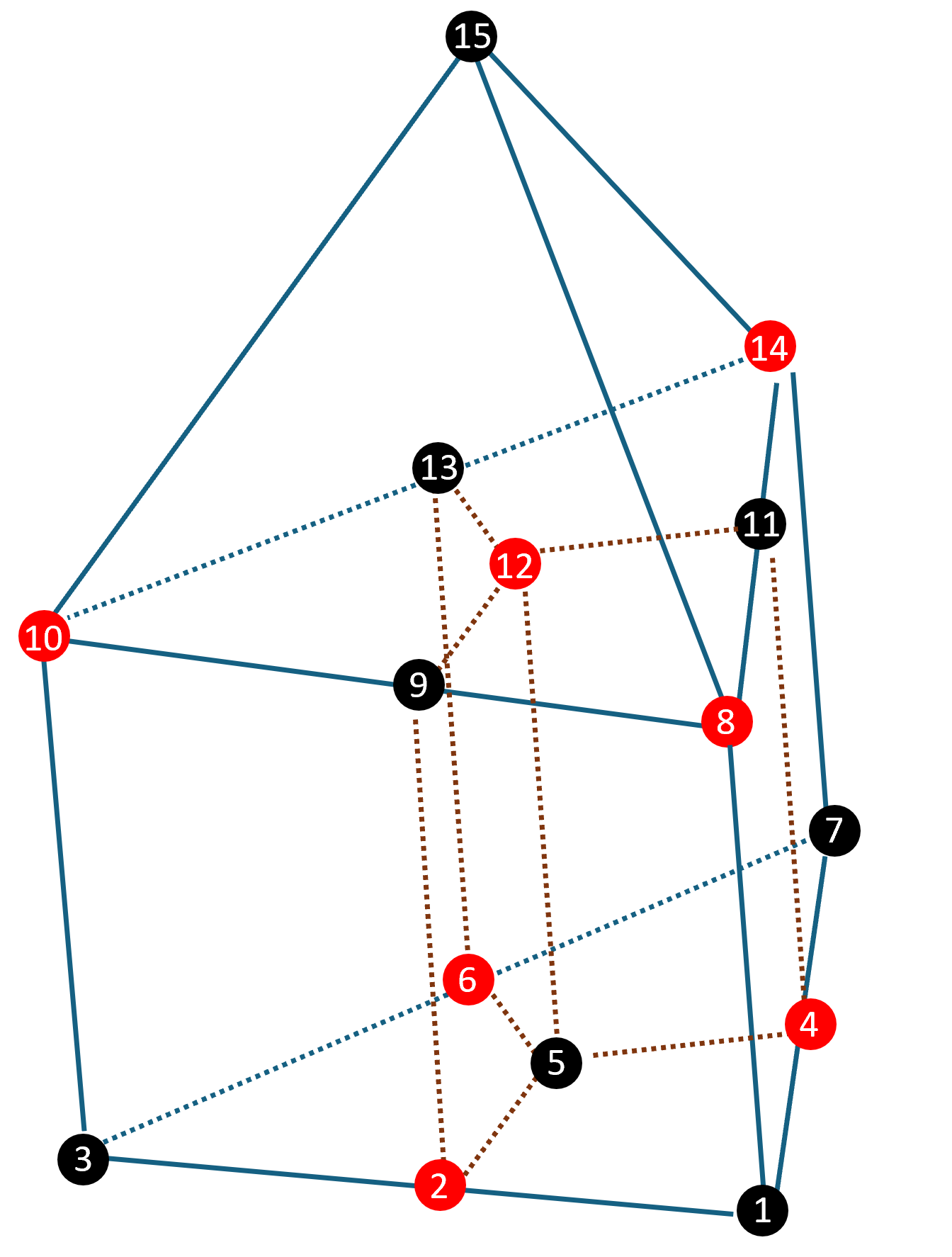}
\caption{\textbf{The visualization of the $3$D [[15,1,3]] color code.} To implement transversal $T$ gate on $3$D color code, one can simultaneously apply $T^\dagger$ gates on all of the black qubits and $T$ gates on all of red qubits.}
\label{fig:15_1_3}
\end{figure}

The $3$D color code [[15,1,3]] is also known as the quantum Reed--Muller (QRM) code. This code can be constructed from the [[7,1,3]] Steane code using the doubling transformation with fixed gauges~\cite{Bravyi2015}. It has four weight-$8$ $X$-type stabilizers and ten weight-$4$ $Z$-type stabilizers, which are associated with the cells and faces, respectively, in Fig.~\ref{fig:15_1_3}. The corresponding $X$ checks are
\begin{align}\label{eq:X_s_10}
 S_{X,[[15,1,3]]}=\begin{pmatrix}
 X_1X_2X_4X_5X_8X_9X_{11}X_{12}\\
 X_2X_3X_5X_6X_9X_{10}X_{12}X_{13}\\
 X_4X_5X_6X_7X_{11}X_{12}X_{13}X_{14}\\
 X_8X_9X_{10}X_{11}X_{12}X_{13}X_{14}X_{15}
 \end{pmatrix}
\end{align}
and the $Z$ checks are
\begin{align}\label{eq:Z_s_10}
 S_{Z,[[15,1,3]]}=\begin{pmatrix}
 Z_1Z_2Z_4Z_5\\
 Z_2Z_3Z_5Z_6\\
 Z_4Z_5Z_6Z_7\\
 Z_8Z_9Z_{11}Z_{12}\\
 Z_9Z_{10}Z_{12}Z_{13}\\
 Z_{11}Z_{12}Z_{13}Z_{14}\\
 Z_{10}Z_{13}Z_{14}Z_{15}\\
 Z_{3}Z_{6}Z_{10}Z_{13}\\
 Z_{2}Z_{3}Z_{9}Z_{10}\\
 Z_{1}Z_{2}Z_{8}Z_{9}
 \end{pmatrix}.
\end{align}

\begin{figure}[t]
\centering
\includegraphics[width=1\linewidth]{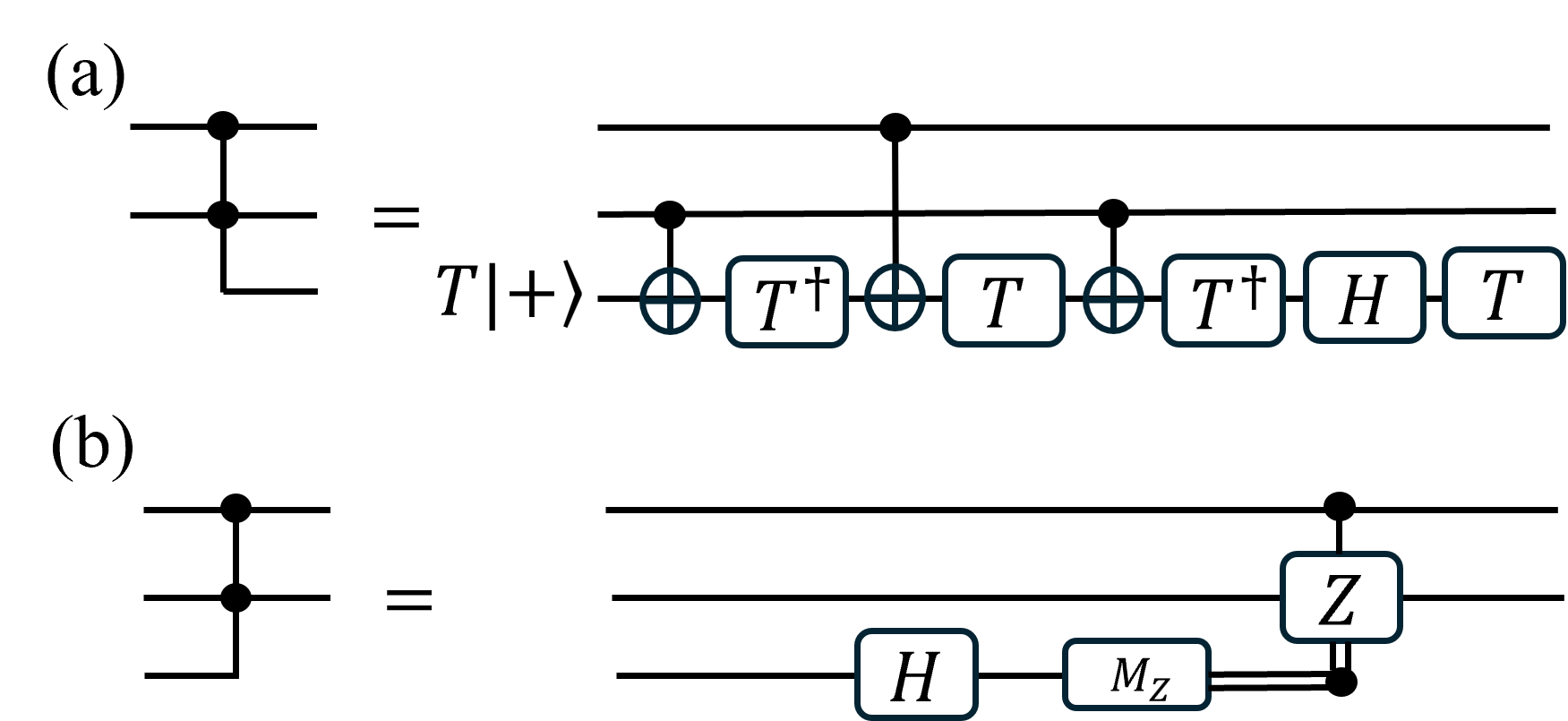}
\caption{\textbf{Circuits for computing (a) and uncomputing (b) logical AND operation.} These operations are necessary for the catalysis process. The circuit in (a) is the costly part since it demands four non-Clifford $T$ gates, while the circuit in (b) only demands Clifford operations and $Z$ measurements. Because of that, most of the cost in the catalysis process comes from computing the logical AND.}
\label{fig:catalyst_2}
\end{figure}

The weight-$3$ $Z$ logical operators are supported on the six edges of the tetrahedron. The corresponding code distance for $Z$-type errors is $d_Z=3$. On the other hand, the weight-$8$ $X$ logical operators are supported on the faces of the tetrahedron. The code distance for $X$-type errors is $d_X=7$.


To construct a $d=5$ doubled color code using the doubling construction, one needs a $d=3$ triorthogonal code with parameters [[15,1,3]] and a $d=5$ self-dual code with parameters [[$n_{sd}$, $1$, $5$]]. The resulting doubled code then has parameters [[$15+2n_{sd}$,$1$,$5$]]. For example, one can choose the [[19,1,5]] ([[17,1,5]]) $2$D color code to construct the $d=5$ doubled color code [[53,1,5]] ([[49,1,5]]). In Fig.~\ref{fig:d_5_3D_code}, based on Ref.~\cite{Tansuwannont2022}, the recursive capped color code provides one such realization. The code consists of a [[15,1,3]] code stacked with two layers of $2$D color codes. The $Z$ stabilizers are associated with the faces, while the $X$ stabilizers are associated with the cells in Fig.~\ref{fig:d_5_3D_code}. Since the [[53,1,5]] ([[49,1,5]]) code has bottom-layer $Z$ stabilizers that coincide with those of the [[19,1,5]] ([[17,1,5]]) $2$D color code, one can switch from the doubled color code to the $2$D color code using a one-way transversal CNOT.

\section{Numerical Simulation}

In this section, we provide details on the state-vector simulation, the two noise models used for the MSC simulations, and additional simulations of ungrown MSC for $d=3$ and $d=5$.

\begin{figure*}[t]
\centering
\includegraphics[width=1\linewidth]{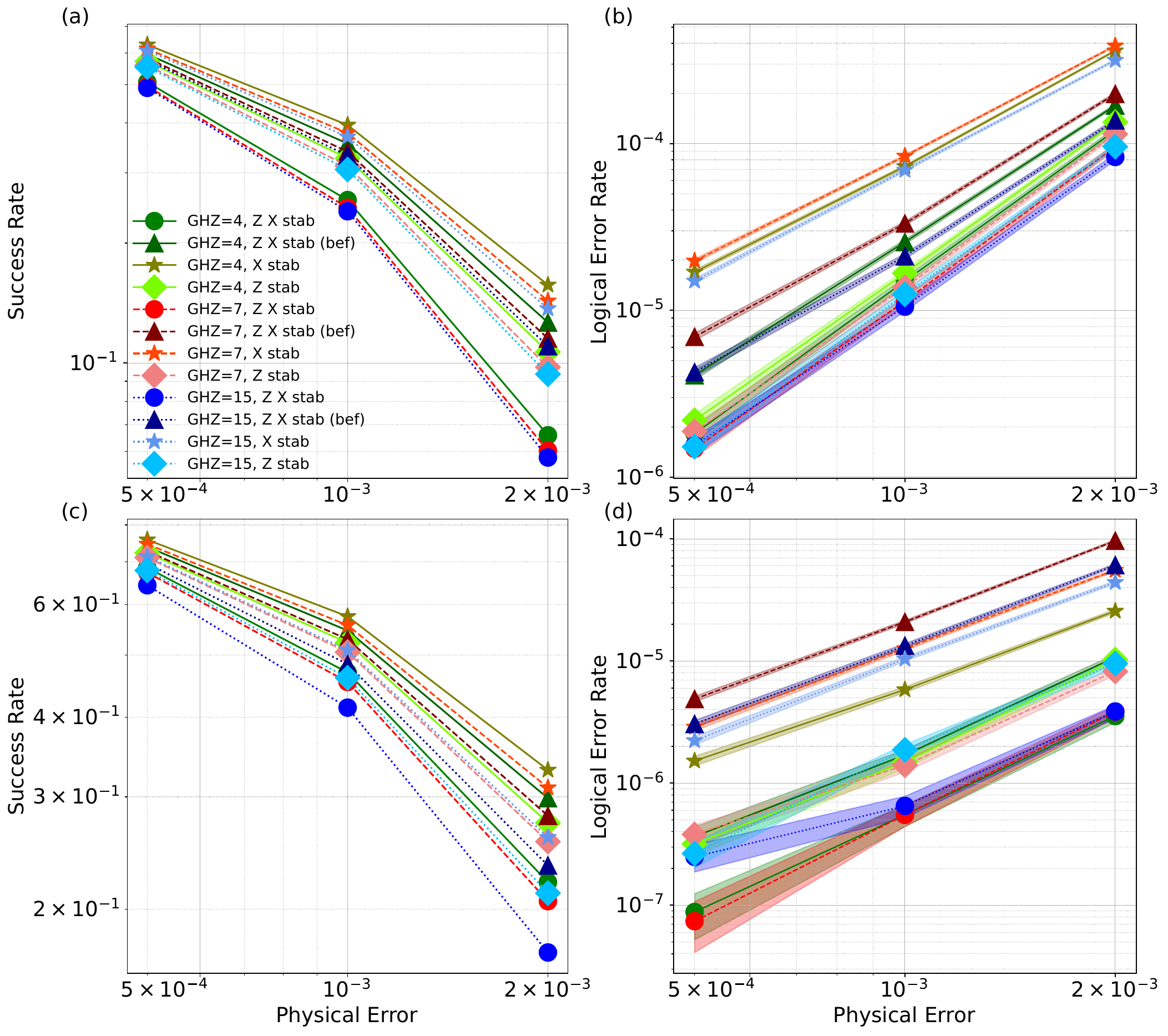}
\caption{
\textbf{The performance of ungrown $d=3$ $\sqrt{T}\ket{+}$ magic-state cultivation using the code-switching method with different GHZ states and syndrome-measurement choices.} The success rate (right panel) and the logical error rate (left panel) vary with the physical noise strength. The upper panels, (a) and (b), show results with idling noise, while the lower panels, (c) and (d), show results without idling noise. Here, GHZ denotes the number of qubits in the GHZ state used. Z Stab (X Stab) denotes the process with only $Z$ (only $X$) stabilizer measurements before and after the double phase-kickback check. Moreover, Z X Stab (Z X Stab (bef)) denotes the process with both $Z$ and $X$ stabilizer measurements before and after (only before) the double phase-kickback check. The shaded regions indicate the standard error obtained from sampling.}
\label{fig:results_ungrown}
\end{figure*}

\subsection{Approximate noisy circuit for state-vector p simulation}

In our state-vector simulation of ungrown MSC on the $[[15,1,3]]$ code, we use Bell states to perform stabilizer measurements. Therefore, the $Z$ stabilizer measurements require $20$ ancilla qubits, since the $[[15,1,3]]$ code has $10$ $Z$ stabilizers. Similarly, the $X$ stabilizer measurements require $8$ ancilla qubits. Thus, the full state-vector simulation would require at least $35$ qubits in total, including both data and ancilla qubits, which makes the simulation computationally challenging.

To reduce this cost, we instead use only two ancilla qubits and measure the stabilizers sequentially, one at a time. For each stabilizer measurement, we apply the corresponding local idling noise to the qubits involved in that measurement. However, this local idling noise still propagates through the circuit during each stabilizer measurement. Consequently, the simulation results obtained from this approximate circuit differ from those obtained using the full circuit. Here, our goal is only to test the consistency of the performance among ungrown MSC protocols for $T\ket{+}_L$, $\sqrt{T}\ket{+}_L$, and $S\ket{+}_L$.

\subsection{Noise Model}

Here, the idling noise model used in the noisy simulation is a uniform noise model with idling noise. That is, after each gate, after qubit initialization, and before measurement, an error occurs with probability $p$. In addition, when qubits are idle, an error also occurs with probability $p$. For initialization and measurement, a bit-flip error (phase-flip error) with probability $p$ is applied after initialization in the $Z$ ($X$) basis or before measurement in the $Z$ ($X$) basis. For a one-qubit gate, including the idling operation, one of the Pauli errors in $\{X,Y,Z\}$ occurs after the gate with probability $p/3$. Similarly, for a two-qubit gate, one of the nontrivial two-qubit Pauli errors in $\{I,X,Y,Z\}^{\otimes 2}/I^{\otimes 2}$ occurs with probability $p/15$.


On the other hand, we also consider a second noise model that does not include idling noise. This is the physics-inspired model considered in Ref.~\cite{Sahay2025}. For nonlocal two-qubit gates, the probability of a two-qubit error is $5p$. In contrast, the error rate for local two-qubit gates remains $p$. Here, we assume that the qubits used for stabilizer measurement are arranged so that the measurements can be performed locally. However, during the double phase-kickback check and the transversal CNOT between the doubled color code and the $2$D color code, the corresponding CNOT gates are nonlocal. In addition, the error rate for Clifford gates is $0.1p$. The initialization and measurement error rates remain $p$.

\subsection{Additional Double Phase Kickback Check and code switch Simulation}


Here, we present additional ungrown MSC simulation results using code switching and different syndrome-extraction choices before and after the double phase-kickback measurement. For the $d=3$ case, in addition to performing a complete syndrome measurement, denoted by $Z$ $X$ Stab, and performing only $Z$ stabilizer measurements, denoted by $Z$ Stab, we also simulate the case with only $X$ stabilizer measurements, denoted by $X$ Stab, and the case with a complete stabilizer measurement only before the double phase-kickback check, denoted by $Z$ $X$ Stab (bef).


Fig.~\ref{fig:results_ungrown} shows the performance of $d=3$ ungrown MSC with different stabilizer choices. In the idling-noise case, although the success rate with only $X$ stabilizer measurements is much higher because fewer measurements are performed, as shown in Fig.~\ref{fig:results_ungrown}(a), the corresponding logical infidelity is much worse than in the other cases, as shown in Fig.~\ref{fig:results_ungrown}(b). The results with both $Z$ and $X$ stabilizer measurements only before the double phase-kickback check show a similar issue. In the case without idling noise, the results with only $X$ stabilizer measurements and those with both $Z$ and $X$ stabilizer measurements only before the double phase-kickback check still perform worse than the other cases in terms of infidelity. These performance discrepancies may arise because the $Z$ syndrome measurements filter out $X$ errors before the one-way transversal CNOT. In particular, the transversal CNOT propagates these $X$ errors to the $2$D color code, thereby inducing logical errors. Hence, the cases without $Z$ stabilizer measurements before the one-way transversal CNOT ultimately have worse infidelity.


Table~\ref{table: ungrown_49all} shows all ungrown $49$-qubit MSC simulation results. Without the second double phase-kickback measurement, the logical infidelities of the MSC protocols using different GHZ states are approximately $3\times10^{-8}$--$8\times10^{-8}$. By adding a second double phase-kickback check without stabilizer measurements between the two checks, the corresponding logical infidelities become approximately $10^{-8}$. With additional $Z$ stabilizer measurements or both $Z$ and $X$ stabilizer measurements between the two checks, the logical infidelities improve by a factor of $10$. On the other hand, without the $X$ stabilizer measurements between or after the two checks, the logical error is also roughly $10^{-8}$. A similar pattern also applies to the $53$-qubit MSC, whose corresponding results are shown in Table~\ref{table: ungrown_53all}.

\section{Lattice Surgery Between Doubled Color Code and Rotated Surface Code}
\label{sec:dcc}
\begin{figure}[t]
\centering
\includegraphics[width=0.7\linewidth]{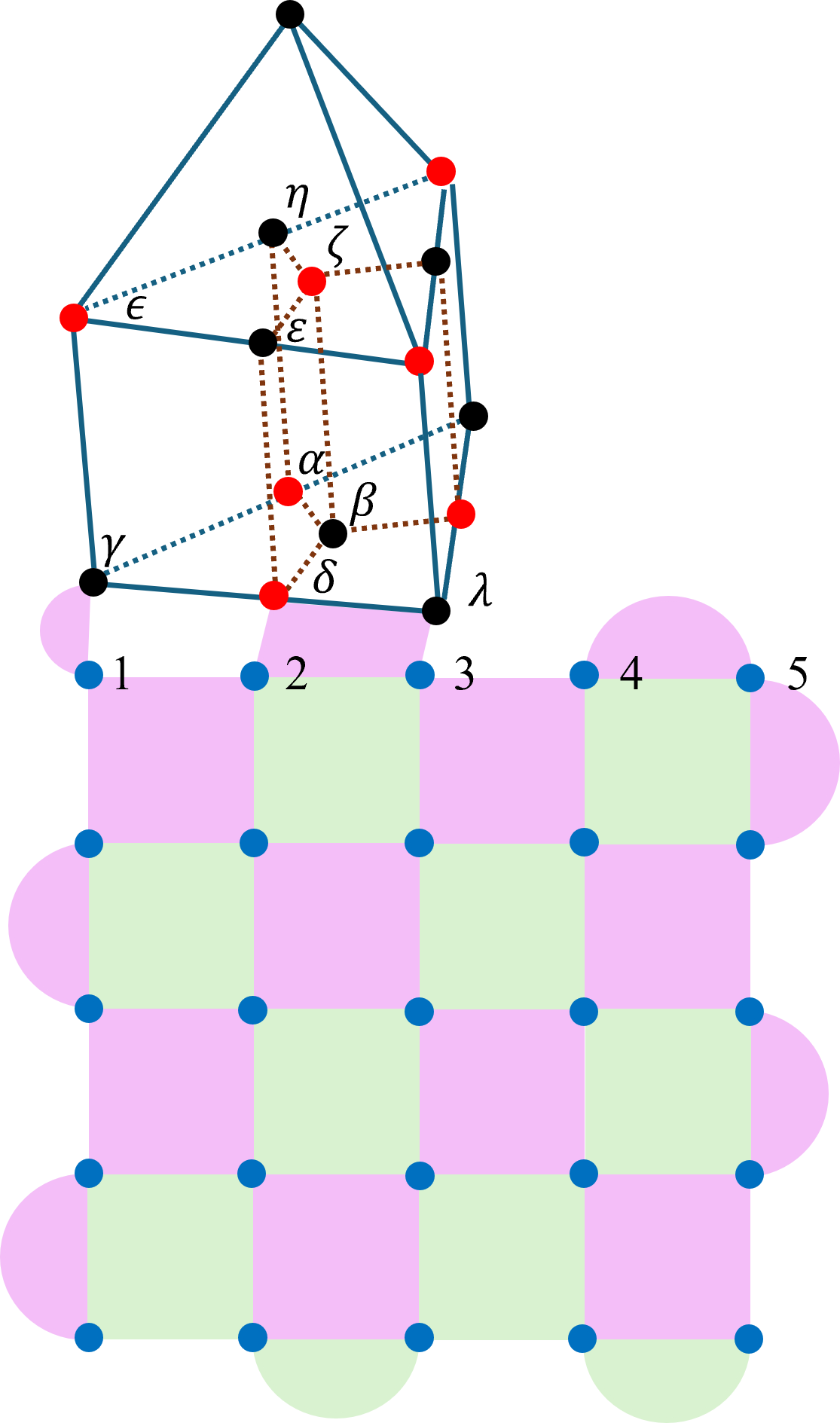}
\caption{\textbf{Lattice surgery between doubled color code and rotated surface code.} 
After the first round of syndrome measurements, the stabilizer $X_{\alpha}X_{\beta}X_{\gamma}X_{\delta}X_{\epsilon}X_{\varepsilon}X_{\zeta}X_{\eta}$ from the doubled color code becomes the new stabilizer $X_{\alpha}X_{\beta}X_{\gamma}X_{\delta}X_{\epsilon}X_{\varepsilon}X_{\zeta}X_{\eta}X_1X_2$. The product of the measurement outcomes for $Z_\gamma Z_1$, $Z_\delta Z_\lambda Z_2 Z_3$, and $Z_4 Z_5$ gives the $ZZ_L$ measurement between the two codes, which determines whether a logical $X$ is applied to the surface code. The pink (green) patches denote the $X$ ($Z$) stabilizers.}  
\label{fig:LS_3D}
\end{figure}


To switch directly from the doubled color code to the rotated surface code, we use a lattice-surgery (LS) procedure inspired by Refs.~\cite{Hirano2025,Chen2026}. Specifically, we perform LS directly between the doubled color code and the surface code. The corresponding LS procedure is as follows.
\begin{enumerate}
    \item All qubits in the rotated surface-code patch are initialized in $\ket{+}$ after the preparation of $T\ket{+}_L$ on the $3$D color code.

    \item The surface-code stabilizers shown in Fig.~\ref{fig:LS_3D} are measured three times. Meanwhile, the $X$ or $Z$ stabilizers of the $3$D color code are measured. During the even-numbered cycles of the surface-code syndrome measurement, the $X$ stabilizers of the color code are measured, except for the stabilizer $X_\alpha X_\beta X_\gamma X_\delta X_\epsilon X_\varepsilon X_\zeta X_\eta$, since it becomes $X_1X_2X_\alpha X_\beta X_\gamma X_\delta X_\epsilon X_\varepsilon X_\zeta X_\eta$ after the first round of syndrome measurement. During the other cycles, the $Z$ stabilizer measurements are performed. Moreover, the boundary $Z$ stabilizers between the color code and the surface code, namely $Z_\gamma Z_1$, $Z_\delta Z_\lambda Z_2 Z_3$, and $Z_4 Z_5$, are measured in every surface-code cycle. The product of these stabilizer outcomes gives the $ZZ_L$ result, which determines whether the $X_L$ operator is applied to the surface code after LS.

    \item All data qubits in the $3$D color code are measured in the $X$ basis. The measurements of the color-code $X$ stabilizers, including $X_1X_2X_\alpha X_\beta X_\gamma X_\delta X_\epsilon X_\varepsilon X_\zeta X_\eta$, can also be obtained at the same time. If any of these measurements gives a nontrivial syndrome outcome, the result is discarded. The $X_L$ measurement is determined by whether the total number of measurement outcomes equal to $1$ is even or odd. If it is even, then $X_L=0_L$; otherwise, $X_L=1_L$. This also determines whether the $Z_L$ operator is applied to the surface code after LS.
\end{enumerate}

According to the procedure above, the $X$ and $Z$ stabilizers from the doubled color code are measured only twice, including the final measurement of the doubled color code in the $X$ basis. Since we postselect on all stabilizer outcomes from the doubled color code and from the boundary between the doubled color code and the surface code, including $Z_4 Z_5$, the procedure remains fault-tolerant.

\section{Spacetime volume estimation}
\label{sec: spt_ve}

 \begin{table*}[]
\begin{tabular}{|c|c|c|}
 \hline
Process & Spacetime Volume & Description \\ \hline \
$[[15,1,3]]$ $\ket{+}_L$ preparation & $23\times8+15\times2$ & 2 4-qubit GHZ states for checking $\ket{+}_L$ \\ \hline \
$[[7,1,3]]$ $\ket{0}_L$ preparation & $7\times2+9\times6$ & a Bell pair for checking $\ket{0}_L$ \\ \hline \
CNOT ladder injection & $15 \times 5$ & Implement noisy $\sqrt{T}$ gate using CNOT ladder \\ \hline \
Z type Stabilizers measurements & $35\times8$ & 10 Bell states for Z-type syndrome extraction \\ \hline \
X type Stabilizers measurements & $23\times8$ & 4 Bell states for X-type syndrome extraction \\ \hline \
$15$q GHZ double phase kickback check &  $13\times30$ & To check $[[15,1,3]]$ magic state  \\ \hline \
$7$q GHZ double phase kickback check & $17\times22$ & To check $[[15,1,3]]$ magic state  \\ \hline \
$4$q GHZ double phase kickback check & $17\times19$ & To check $[[15,1,3]]$ magic state  \\ \hline \
Transversal T & 15  & Apply $T$ and $T^\dagger$ on each physical qubit of [[15,1,3]] code \\ \hline \
Logical $X$ measurement on $[[15,1,3]]$ & 15  & Measure each physical qubit in $X$ basis \\ \hline \
Transversal CNOT & 22  & Transversal CNOT between Steane code and doubled color code \\ \hline \
$2$D Lattice Surgery ($d_i=3$ to $d_f=11$) & $255\times21+241\times6$   & $3$ rounds of surface code syndrome measurement \\ \hline \
$3$D Lattice Surgery ($d_i=3$ to $d_f=11$) & $277\times24+241\times6$   & $3$ rounds of surface code syndrome measurement\\ \hline \
$10$ rounds of syndrome measurement & $241\times6\times10$   & Waiting for complementary gap calculation \\ \hline
\end{tabular}
\caption{\textbf{The spacetime volume of different parts of magic state preparation. 
} 
For $3$D LS, immediately after the double phase-kickback check, lattice surgery is performed directly to connect the doubled color code with $d=3$ to the rotated surface code with $d_f=11$. In contrast, for $2$D LS, after the double phase-kickback check, one can perform $Z$ and $X$ stabilizer measurements, or use other stabilizer-measurement choices, before applying the one-way transversal CNOT.}
\label{tab: spt}
\end{table*}


Here, we provide more details on the space-time volume estimation for each part of the MSC protocol. We also consider how different parts of the circuit are connected and parallelize parts of the circuit so that the total space-time volume is optimized. However, for the $d=5$ case, the corresponding space-time volume is too large because we do not optimize the circuit depth of the stabilizer measurements or the parallelization between the doubled color-code and surface-code cycles. We leave this optimization to future work.

Table~\ref{tab: spt} provides the optimized space-time volume for each part of the circuit. The protocol begins with the initialization of the $[[15,1,3]]$ code in $\ket{+}_L$. The initialization circuit has depth $5$, while the circuit for checking the quality of $\ket{+}_L$ using two $4$-qubit GHZ states has depth $8$. However, the preparation of the two $4$-qubit GHZ states has circuit depth $3$ and can be parallelized with part of the $[[15,1,3]]$ $\ket{+}_L$ initialization. Similarly, for the $[[7,1,3]]$ $\ket{0}_L$ initialization, the preparation of a Bell pair can be parallelized with part of the $[[7,1,3]]$ $\ket{+}_L$ preparation.

For the $Z$ stabilizer measurements, we use $10$ Bell states to measure all stabilizers simultaneously. Similarly, we use $4$ Bell states to obtain the $X$ syndrome measurements. For the double phase-kickback check, as shown in Fig.~\ref{fig:dp_ST}(b), one can pair each controlled-$XT$ gate with its adjoint controlled-$XT^\dagger$ gate on the same control qubit from the GHZ state to cancel the extra phase; this requires a $7$-qubit GHZ state. Similarly, one can use two controlled-$XT$ gates and two controlled-$XT^\dagger$ gates to cancel the phase with a $4$-qubit GHZ state.

For the $2$D ($3$D) LS procedure, the circuit depth of the $X$ or $Z$ stabilizer measurements is one (two) more than that of a surface-code syndrome cycle. Thus, the total circuit depth of one round of LS is $7$ ($8$). After all $2$D color-code (doubled color-code) qubits are measured, one extra round of surface-code syndrome measurement is performed. After LS, $10$ rounds of surface-code syndrome measurements are applied while waiting for the decoder to compute the complementary gap.


For the expected space-time volume per kept shot, $V$ in eq.~\ref{eq:stv}, we also compute the corresponding success rates for different stages of the MSC circuit. Moreover, we select reasonable target logical infidelities for MSC with different escape strategies and stabilizer-measurement choices to evaluate the corresponding success rates after growth.

\section{Catalyze $\sqrt{T}$ gates and magic state teleportation}


According to Ref.~\cite{Bravyi2005}, it is possible to implement the $\sqrt{T}$ gate using only the preparation of $\ket{0}_L$, Clifford operations, measurement in the $Z$ basis, and the $\sqrt{T}\ket{+}_L$ magic state. However, this method requires a large number of copies of $\sqrt{T}\ket{+}_L$, on average, for each implementation of $\sqrt{T}$. But, if the system is allowed to implement $T$ gates, a $\sqrt{T}$ gate can instead be applied using only one $T$ gate and one $\sqrt{T}\ket{+}_L$ magic state through the magic-state teleportation (MST) circuit shown in Fig.~\ref{fig:mst}. The $T$ gate can be implemented using a similar circuit by replacing the $T$ gate with an $S$ gate and replacing $\sqrt{T}\ket{+}_L$ with $T\ket{+}_L$. Thus, the only non-Clifford resources required to implement the $\sqrt{T}$ gate are one copy of each of the magic states $\sqrt{T}\ket{+}_L$ and $T\ket{+}_L$. Since the application of the $T$ gate is conditioned on a measurement result that occurs with probability $50\%$, in practical algorithms requiring many $\sqrt{T}$ gates only half of the MST applications consume a $T$ gate on average.


The preparation of $\sqrt{T}\ket{+}_L$ using current techniques is very costly, making MST prohibitive for algorithms that require many $\sqrt{T}$ gates. To bypass this problem, magic-state catalysis (MSCa) methods were developed. These methods use a circuit with one $\sqrt{T}\ket{+}$ state, referred to as the seed state, to ``transform'' a $T$ gate into two applications of $\sqrt{T}$. Additional $T$ gates may also be needed to construct the catalysis circuit, so the non-Clifford cost of catalysis is greater than just one $T$ gate for producing two $\sqrt{T}$ gates. The advantage of these methods is that they only ``spend'' the $T$ gates, allowing the seed state to be reused in subsequent catalysis procedures. To our knowledge, the best catalysis method is presented in Fig.~\ref{fig:catalyst}, with the computation and uncomputation of the logical AND shown in Fig.~\ref{fig:catalyst_2}(a) and (b), respectively. The application of a $\sqrt{T}^3$ gate can also be performed using MSCa by replacing the seed with the $\sqrt{T}^3\ket{+}_L$ state and adding an $S$ gate after the $T$ gate in the middle of the circuit.


Given these considerations, an MSC protocol capable of producing $\sqrt{T}\ket{+}_L$, such as the one proposed in this paper, can make MST a viable method for producing $\sqrt{T}$ gates at lower cost than MSCa, as discussed in the main text in Sec.~\ref{sec:FFTQC}.

\begin{table*}[]
\begin{tabular}{|c|c|c|}
 \hline
             & Z & X \\ \hline
$[[49,1,5]]$ & $\begin{pmatrix}
 Z_2Z_3Z_5Z_6\\
 Z_3Z_4Z_6Z_7\\
 Z_5Z_6Z_7Z_8\\
 Z_9Z_{10}Z_{12}Z_{14}\\
  Z_{10}Z_{11}Z_{13}Z_{14}\\
  Z_{11}Z_{12}Z_{13}Z_{14}\\
  Z_1Z_2Z_5Z_8\\
  Z_2Z_5Z_9Z_{12}\\
  Z_2Z_3Z_9Z_{10}\\
  Z_3Z_4Z_{10}Z_{11}\\
  Z_{16}Z_{17}Z_{18}Z_{19}\\
  Z_{33}Z_{34}Z_{35}Z_{36}\\
  Z_{17}Z_{19}Z_{22}Z_{23}\\
  Z_{34}Z_{36}Z_{39}Z_{40}\\
  Z_{22}Z_{23}Z_{26}Z_{27}\\
  Z_{39}Z_{40}Z_{43}Z_{44}\\
  Z_{26}Z_{27}Z_{31}Z_{32}\\
  Z_{43}Z_{44}Z_{48}Z_{49}\\
  Z_{18}Z_{19}Z_{22}Z_{26}Z_{31}Z_{30}Z_{25}Z_{21}\\
  Z_{35}Z_{36}Z_{39}Z_{43}Z_{48}Z_{47}Z_{42}Z_{38}\\
  Z_{20}Z_{21}Z_{24}Z_{25}\\
  Z_{37}Z_{38}Z_{41}Z_{42}\\
  Z_{24}Z_{25}Z_{29}Z_{30}\\
  Z_{41}Z_{42}Z_{46}Z_{47}\\
  Z_{20}Z_{24}Z_{28}Z_{29}\\
  Z_{37}Z_{41}Z_{45}Z_{46}\\
  Z_{9}Z_{12}Z_{15}Z_{28}Z_{29}Z_{30}Z_{31}Z_{32}\\
  Z_{20}Z_{21}Z_{37}Z_{38}\\
  Z_{20}Z_{28}Z_{37}Z_{45}\\
  Z_{18}Z_{19}Z_{35}Z_{36}\\
  Z_{19}Z_{22}Z_{36}Z_{39}\\
  Z_{22}Z_{26}Z_{39}Z_{43}\\
  Z_{26}Z_{31}Z_{43}Z_{48}\\
  Z_{31}Z_{32}Z_{48}Z_{49}\\
  Z_{28}Z_{29}Z_{45}Z_{46}
 \end{pmatrix}$
&  $\begin{pmatrix}
 X_1X_2X_3X_4X_5X_6X_7X_8\\
 X_2X_3X_5X_6X_9X_{10}X_{12}X_{14}\\
 X_3X_4X_6X_7X_{10}X_{11}X_{14}X_{13}\\
 X_5X_6X_7X_8X_{11}X_{14}X_{13}X_{15}\\
 S_{24}\\
 X_{16}X_{17}X_{18}X_{19}X_{33}X_{34}X_{35}X_{36}\\
 X_{17}X_{19}X_{22}X_{23}X_{34}X_{36}X_{39}X_{40}\\
 X_{22}X_{23}X_{26}X_{27}X_{39}X_{40}X_{43}X_{44}\\
 X_{26}X_{27}X_{31}X_{32}X_{43}X_{44}X_{48}X_{49}\\
 X_{18}X_{19}X_{21}X_{26}X_{31}X_{30}X_{25}X_{21}X_{35}X_{36}X_{39}X_{43}X_{48}X_{47}X_{42}X_{38}\\
 X_{20}X_{21}X_{24}X_{25}X_{37}X_{38}X_{41}X_{42}\\
 X_{24}X_{25}X_{29}X_{30}X_{41}X_{42}X_{46}X_{47}\\
 X_{20}X_{24}X_{28}X_{29}X_{37}X_{41}X_{45}X_{46}
 \end{pmatrix}$ \\ \hline
\end{tabular}
\caption{\textbf{The stabilizers of $[[49,1,5]]$. $S_{24}=X_9X_{10}X_{11}X_{12}X_{14}X_{13}X_{15}X_{16}X_{17}X_{18}X_{19}
X_{20}X_{21}X_{22}X_{23}X_{24}X_{25}X_{26}X_{27}X_{28}\\X_{29}X_{30}X_{31}X_{32}$} The labels correspond to the label in Fig.~\ref{fig:d_5_3D_code} (b).}
\end{table*}

\begin{table*}[]
\begin{tabular}{|c|c|c|}
 \hline
             & Z & X \\ \hline
$[[53,1,5]]$ & $\begin{pmatrix}
 Z_2Z_3Z_5Z_6\\
 Z_3Z_4Z_6Z_7\\
 Z_5Z_6Z_7Z_8\\
 Z_9Z_{10}Z_{12}Z_{14}\\
  Z_{10}Z_{11}Z_{13}Z_{14}\\
  Z_{11}Z_{12}Z_{13}Z_{14}\\
  Z_1Z_2Z_5Z_8\\
  Z_2Z_5Z_9Z_{12}\\
  Z_2Z_3Z_9Z_{10}\\
  Z_3Z_4Z_{10}Z_{11}\\
  Z_{16}Z_{17}Z_{25}Z_{21}\\
  Z_{35}Z_{36}Z_{44}Z_{40}\\
  Z_{17}Z_{18}Z_{21}Z_{22}\\
  Z_{36}Z_{37}Z_{40}Z_{41}\\
  Z_{18}Z_{19}Z_{23}Z_{27}Z_{26}Z_{22}\\
  Z_{37}Z_{38}Z_{42}Z_{46}Z_{45}Z_{41}\\
  Z_{19}Z_{20}Z_{23}Z_{24}\\
  Z_{38}Z_{39}Z_{42}Z_{43}\\
  Z_{21}Z_{22}Z_{26}Z_{29}Z_{28}Z_{25}\\
  Z_{40}Z_{41}Z_{45}Z_{48}Z_{47}Z_{44}\\
  Z_{23}Z_{24}Z_{30}Z_{27}\\
  Z_{42}Z_{43}Z_{49}Z_{46}\\
  Z_{31}Z_{32}Z_{30}Z_{27}Z_{26}Z_{29}\\
  Z_{50}Z_{51}Z_{49}Z_{46}Z_{45}Z_{48}\\
  Z_{22}Z_{26}Z_{30}Z_{31}\\
  Z_{32}Z_{31}Z_{28}Z_{29}\\
  Z_{34}Z_{32}Z_{31}Z_{33}\\
  Z_{53}Z_{52}Z_{50}Z_{51}\\
  Z_{9}Z_{10}Z_{11}Z_{16}Z_{17}Z_{18}Z_{19}Z_{20}\\
  Z_{19}Z_{20}Z_{38}Z_{39}\\
  Z_{16}Z_{17}Z_{35}Z_{36}\\
  Z_{16}Z_{25}Z_{35}Z_{44}\\
  Z_{21}Z_{22}Z_{40}Z_{41}\\
  Z_{22}Z_{36}Z_{41}Z_{45}\\
  Z_{26}Z_{27}Z_{45}Z_{46}\\
  Z_{29}Z_{26}Z_{45}Z_{48}\\
  Z_{29}Z_{31}Z_{48}Z_{50}\\
  Z_{23}Z_{27}Z_{42}Z_{46}
 \end{pmatrix}$
&  $\begin{pmatrix}
 X_1X_2X_3X_4X_5X_6X_7X_8\\
 X_2X_3X_5X_6X_9X_{10}X_{12}X_{14}\\
 X_3X_4X_6X_7X_{10}X_{11}X_{14}X_{13}\\
 X_5X_6X_7X_8X_{11}X_{14}X_{13}X_{15}\\
 S_{26}\\
 X_{16}X_{17}X_{25}X_{21}X_{35}X_{36}X_{44}X_{40}\\
 X_{17}X_{18}X_{21}X_{22}X_{36}X_{37}X_{40}X_{41}\\
 X_{18}X_{19}X_{23}X_{27}X_{26}X_{22}X_{37}X_{38}X_{42}X_{46}X_{45}X_{41}\\
 X_{19}X_{20}X_{23}X_{24}X_{38}X_{39}X_{42}X_{43}\\
 X_{21}X_{22}X_{26}X_{29}X_{28}X_{25}X_{40}X_{41}X_{45}X_{48}X_{47}X_{44}\\
 X_{23}X_{24}X_{30}X_{27}X_{42}X_{43}X_{49}X_{46}\\
 X_{31}X_{32}X_{30}X_{27}X_{26}X_{29}X_{50}X_{51}X_{49}X_{46}X_{45}X_{48}\\
 X_{52}X_{50}X_{47}X_{48}X_{33}X_{31}X_{28}X_{29}\\
 X_{34}X_{32}X_{31}X_{33}X_{53}X_{52}X_{50}X_{51}
 \end{pmatrix}$ \\ \hline
\end{tabular}
\caption{\textbf{The stabilizers of $[[53,1,5]]$.} $S_{26}=X_9X_{10}X_{11}X_{12}X_{14}X_{13}X_{15}X_{16}X_{17}X_{18}X_{19}
X_{20}X_{21}X_{22}X_{23}X_{24}X_{25}X_{26}X_{27}X_{28}\\X_{29}X_{30}X_{31}X_{32}X_{33}X_{34}$ is a $26$-weight stabilizer. The labels correspond to the label in Fig.~\ref{fig:d_5_3D_code} (a).} 
\end{table*}

\section{The STAR Architecture}

\begin{figure}[t]
\centering
\includegraphics[width=0.7\linewidth]{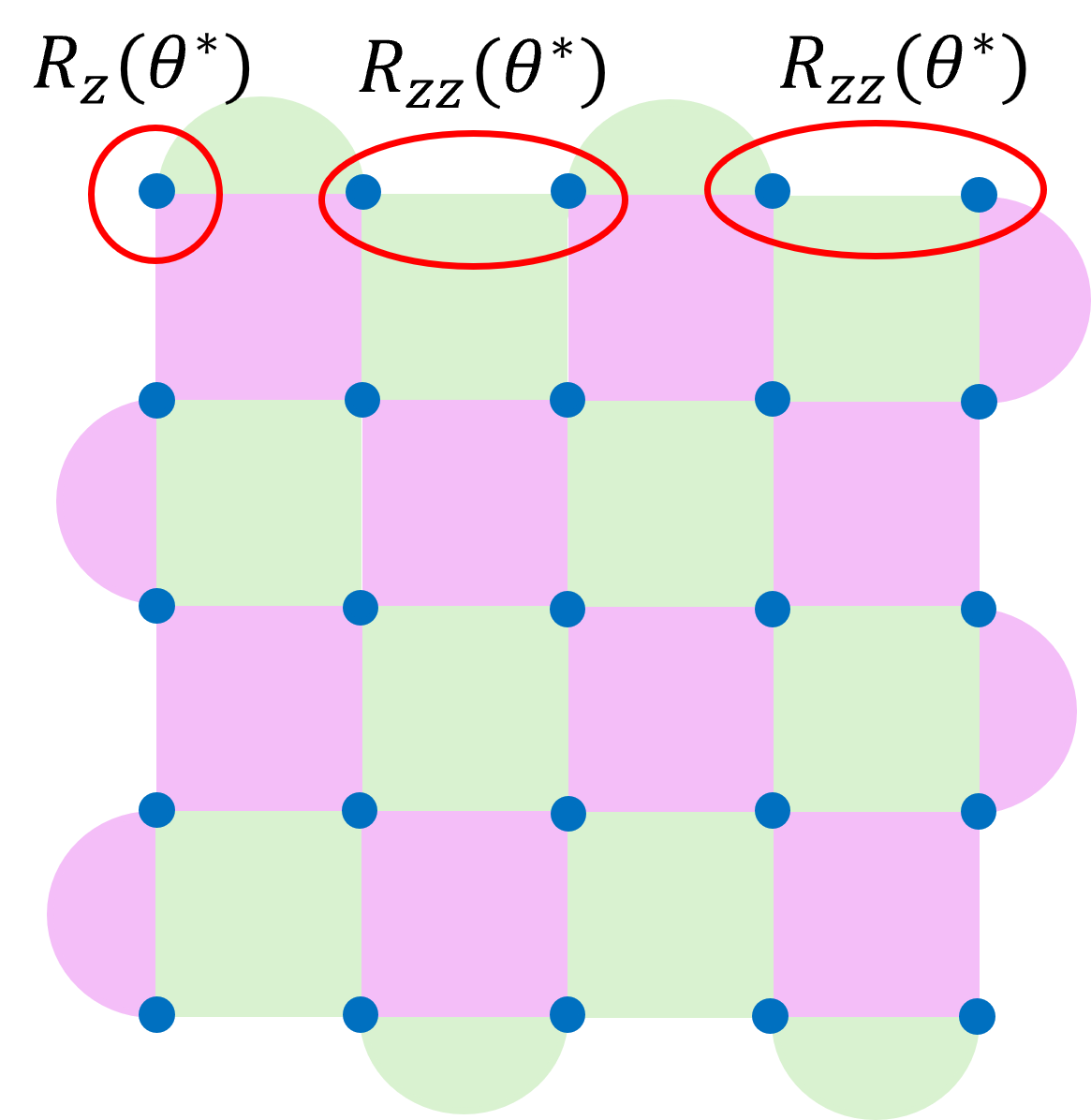}
\caption{
\textbf{Example of transversal multi-qubit rotation on the $d=5$ rotated surface code.} A single-qubit $Z$ rotation gate and two two-qubit $Z$ rotation gates, indicated by red circles, are applied with physical rotation angle $\theta^*$ to the qubits that support the logical $Z$ gate. The pink (green) patches denote the $X$ ($Z$) stabilizers.}
\label{fig:tmr}
\end{figure}
The STAR architecture is designed for quantum simulation, which usually requires the gate set Clifford+$R_z(\theta_r)$ with a small rotation angle $\theta_r$. The transversal multi-qubit rotation (TMR) protocol provides access to small-angle rotations at low cost and with low noise. In this protocol, one initializes the logical state $\ket{+}_L$ and applies $k$ multi-qubit $Z$ rotations to the physical qubits that support the logical $Z$ operator. For instance, in Fig.~\ref{fig:tmr}, $k=3$ $Z$ rotation gates are applied along the edge of the surface code that supports the logical $Z$ operation. After these $k$ multi-qubit rotations, the state becomes
\begin{align}
\begin{split}
\label{eq:R_z}
    &\quad\prod_{i}^{k} R_{Z_s,i}(\theta^*)\ket{+}_L\\
   & =\cos^k\left(\frac{\theta^*}{2}\right)\ket{+}_L +(-i)^k \sin^k\left(\frac{\theta^*}{2}\right)\ket{-}_L\\
   &+(\text{$Z$-error terms})\\&
    = {\sqrt{P_{\text{ideal}}}} \left [ \cos(\frac{\theta_r}{2})\ket{+}_L +(-i)^k \sin(\frac{\theta_r}{2})\ket{-}_L \right ]+\\&
    (\text{$Z$-error terms}) \; ,
\end{split}
\end{align}
where $R_{Z_s,i}$ denotes a single- or multi-qubit $Z$ rotation, with $Z_s$ denoting a Pauli $Z$ string operator. The parameter $\theta^*$ is the physical rotation angle and satisfies
$
\frac{\theta_r}{2}
=\arcsin\left(
\frac{\sin^k(\theta^*/2)}
{\sqrt{\sin^{2k}(\theta^*/2)+\cos^{2k}(\theta^*/2)}}
\right),
$
and
$
P_{\text{ideal}}
=
\cos^{2k}\left(\frac{\theta^*}{2}\right)
+
\sin^{2k}\left(\frac{\theta^*}{2}\right)
$
is the ideal success probability in the absence of noise. The last term represents the component of the state outside the code space, which arises because the multi-transversal rotation gate does not belong to the actual transversal gate set of the target code. To eliminate these terms, one can perform two rounds of syndrome measurement and fully or partially postselect on the outcomes. Here, partial postselection means postselecting only on the stabilizer outcomes associated with the qubits on which the TMR operation is applied. Moreover, according to Ref.~\cite{ismail2026transversal}, $X$ errors that occur on qubits involved in the TMR operation before the TMR operation is applied can induce coherent logical errors after error correction, since $X$ errors anticommute with logical $Z$ operators. In addition, readout errors before the TMR operation can also cause large logical errors. Therefore, before the TMR operation, one should also fully or partially postselect on syndrome-measurement outcomes to filter out these dangerous errors.

\begin{table*}[]
\begin{tabular}{|c|c|c|c|c|}
\hline
Tests                                     & GHZ & Success Rate ($\%$) & Infidelity ($10^{-9}$) & Std. Dev. ($10^{-9}$) \\ \hline
w/o mid and sec check                           & 49  & 3.318               & 80.8                   & 7.0                   \\ \hline
w/o mid                                   & 49  & 1.167               & 17.1                   & 5.4                   \\ \hline
w/o $X_{\text{mid}}$ and $X_{\text{aft}}$ & 49  & 1.089               & 16.5                   & 5.5                   \\ \hline
w/o $X_{\text{mid}}$                             & 49  & 0.8112              & 4.9                    & 2.5                   \\ \hline
w/o $X_{\text{aft}}$                             & 49  & 0.8168              & 1.2                    & 1.2                   \\ \hline
All Stab                                  & 49  & 0.6085              & $\lesssim 1.64$                & N/A               \\ \hline
w/o mid and sec check                           & 25  & 4.386               & 52.4                   & 4.9                   \\ \hline
w/o mid                                   & 25  & 2.039               & 6.9                    & 2.6                   \\ \hline
w/o $X_{\text{mid}}$ and $X_{\text{aft}}$ & 25  & 1.901               & 17.9                   & 4.3                   \\ \hline
w/o $X_{\text{mid}}$                             & 25  & 1.417               & 3.5                    & 1.6                   \\ \hline
w/o $X_{\text{aft}}$                             & 25  & 1.427               & 0.70                   & 0.70                  \\ \hline
All Stab                                  & 25  & 1.063               & $\lesssim 0.94$                & N/A               \\ \hline
w/o mid and sec check                           & 12  & 5.083               & 36.6                   & 3.8                   \\ \hline
w/o mid                                   & 12  & 2.739               & 8.8                    & 2.5                   \\ \hline
w/o $X_{\text{mid}}$ and $X_{\text{aft}}$ & 12  & 2.554               & 15.6                   & 3.5                   \\ \hline
w/o $X_{\text{mid}}$                             & 12  & 1.903               & 6.3                    & 1.8                   \\ \hline
w/o $X_{\text{aft}}$                             & 12  & 1.916               & 1.04                   & 0.74                  \\ \hline
All Stab                                  & 12  & 1.428               & $\lesssim 0.7$                 & N/A                \\ \hline
\end{tabular}
\caption{\textbf{The performance of $[[49,1,5]]$ ungrown magic state cultivation with two double phase kickback checks.} mid refers the both $X$ and $Z$ type stabilizer measurements between two double checks. sec check means the second double phase kickback check. $X_{\text{mid}}$ means $X$ type stabilizer measurements in the middle of two double phase kickback checks, while $X_{\text{aft}}$ means $X$ type stabilizer measurements after the second double phase kickback check. Here, we have a total of $10^{11}$ shots for each case. However, the total shot is not enough to estimate the logical infidelities of some cases. For those cases that does not show any logical errors, we estimate the corresponding logical error rate $\lesssim 1/N_\mathrm{eff}$, where $N_\mathrm{eff}$ represents the total number of successful shots. The corresponding standard errors are not applicable. }
\label{table: ungrown_49all}
\end{table*}

\begin{table*}[]
\begin{tabular}{|c|c|c|c|c|}
\hline
Tests                                     & GHZ & Success Rate ($\%$) & Infidelity ($10^{-9}$) & Std. Dev. ($10^{-9}$) \\ \hline
w/o mid and sec                           & 53  & 2.677               & 90.0                   & 5.8                   \\ \hline
w/o mid                                   & 53  & 0.8642              & 19.7                   & 4.8                   \\ \hline
w/o $X_{\text{mid}}$ and $X_{\text{aft}}$ & 53  & 0.8102              & 27.2                   & 5.8                   \\ \hline
w/o $X_{mid}$                             & 53  & 0.5901              & 11.9                   & 4.5                   \\ \hline
w/o $X_{aft}$                             & 53  & 0.5900              & $\lesssim 1.69$                & N/A              \\ \hline
All Stab                                  & 53  & 0.4297              & $\lesssim 2.33$                & N/A               \\ \hline
w/o mid and sec                           & 27  & 3.621               & 50.5                   & 3.7                   \\ \hline
w/o mid                                   & 27  & 1.581               & 8.9                    & 2.4                   \\ \hline
w/o $X_{\text{mid}}$ and $X_{\text{aft}}$ & 27  & 1.482               & 19.6                   & 3.6                   \\ \hline
w/o $X_{mid}$                             & 27  & 1.080               & 4.6                    & 2.1                   \\ \hline
w/o $X_{aft}$                             & 27  & 1.080               & $\lesssim 0.93$                 & N/A           \\ \hline
All Stab                                  & 27  & 0.7862              & 1.3                    & 1.3                   \\ \hline
w/o mid and sec                           & 13  & 4.245               & 47.8                   & 3.4                   \\ \hline
w/o mid                                   & 13  & 2.172               & 10.1                   & 2.2                   \\ \hline
w/o $X_{\text{mid}}$ and $X_{\text{aft}}$ & 13  & 2.037               & 13.3                   & 2.6                   \\ \hline
w/o $X_{mid}$                             & 13  & 1.483               & 2.0                    & 1.2                   \\ \hline
w/o $X_{aft}$                             & 13  & 1.483               & 0.67                & N/A               \\ \hline
All Stab                                  & 13  & 1.080               & 0.93                   & 0.93                  \\ \hline
\end{tabular}
\caption{\textbf{The performance of $[[53,1,5]]$ ungrown magic state cultivation with two double phase kickback checks.} (See caption of table~\ref{table: ungrown_49all})}
\label{table: ungrown_53all}
\end{table*}

However, noise that occurs during the TMR process can combine with these $Z$-error terms and become a coherent logical error. Therefore, the corresponding method is not fault-tolerant. The induced logical error $\epsilon_r$ scales as $O(p_{\mathrm{ph}}\theta_r^{2(1-1/k)})$, where $p_{\mathrm{ph}}$ is the physical qubit error rate~\cite{Toshio2024}. This is the logical error associated with preparing a small-angle magic state, $\ket{m_\theta}_L=R_{Z_L}(\theta)\ket{+}_L$. For the full implementation of the $R_{Z_L}$ gate, one must also account for logical errors induced by teleportation and the RUS protocol. The total logical error is then
\begin{align}
\label{eq: star_el}
     \epsilon_{L,R}\left(\theta_r=\frac{\pi}{2^{n}}\right)
     &=\frac{1}{2^{n-1}}\epsilon_{L,S}
     +\sum_{m=0}^{n-2}\frac{m}{2^m}\epsilon_{L,\text{CNOT}} \notag\\
     &\quad+\sum_{m=0}^{n-2} \frac{1}{2^m}
     \epsilon_r\left(\theta=\frac{\pi}{2^{n-m}}\right).
\end{align}
Here, $\epsilon_{L,G}$ is the logical error associated with implementing the Clifford gate $G$, with $G\in \{S,\text{CNOT}\}$. For certain QEC codes, Clifford gates can be implemented fault-tolerantly using transversal gates or lattice surgery. Once the physical gate error rate is below threshold, the logical error rates of Clifford gates decrease as the code distance increases. Thus, for large QEC codes, the dominant contribution to the logical error of the STAR rotation is the last term in Eq.~\ref{eq: star_el}. This last term is estimated to be approximately $\alpha p_{\mathrm{ph}}|\theta_r|$, with $\alpha$ empirically given by $1.5k$. If one also uses the $k$-switch protocol, the coefficient becomes $\alpha \approx 1.5$.

\vfill\eject

\bibliography{refs}
\end{document}